\documentclass[twocolumn, trackchanges]{aastex631}
\usepackage{lmodern}
\usepackage{amsmath}
\usepackage{graphicx}   
\usepackage{bm}         
\usepackage{natbib}
\usepackage{etoolbox}
\usepackage{rotating}
\usepackage{array}
\usepackage{booktabs}
\usepackage{amssymb}
\usepackage{textcomp}
\usepackage{gensymb}
\usepackage{hyperref}
\usepackage{wrapfig}
\usepackage{ulem}
\usepackage{booktabs}
\usepackage{etoolbox}
\usepackage{lipsum}
\usepackage{enumitem}
\setlist[enumerate]{itemsep=0mm}
\usepackage[table,x11names]{xcolor}

\newcommand{\OIII}{\mbox{[O~\textsc{iii}]}}
\newcommand{\OI}{\mbox{[O~\textsc{i}]}}
\newcommand{\HII}{\mbox{H~\textsc{ii}}}

\newcommand{\NeII}{\mbox{[Ne~\textsc{ii}]}}
\newcommand{\kms}{\mbox{km s$^{-1}$}}

\begin{document}

\title{The SOMA Atomic Outflow Survey. I.\\An Atomic OI and Highly Ionized OIII Outflow from Massive Protostar G11.94-00.62}

\author[0009-0002-9291-1487]{Phillip Oakey}
\affiliation{Department of Astronomy, University of Virginia, Charlottesville, VA, 22904, USA}

\author[0000-0001-8227-2816]{Yao-Lun Yang}
\affiliation{Star and Planet Formation Laboratory, RIKEN Pioneering Research Institute, Wako-shi, Saitama, 351-0106, Japan}

\author[0000-0002-3389-9142]{Jonathan C. Tan}
\affiliation{Department of Astronomy, University of Virginia, Charlottesville, VA, 22904, USA}
\affiliation{Department of Space, Earth, and Environment, Chalmers University of Technology, Gothenburg, Västra Götaland, 412 96, Sweden}

\author[0000-0003-2733-4580]{Thomas G. Bisbas}
\affiliation{Research Center for Astronomical Computing, Zhejiang Lab, Hangzhou, Zhejiang Province, 311121 P.R. China} 

\author[0000-0003-4040-4934]{Rub{\'e}n Fedriani}
\affiliation{Instituto de Astrof\'isica de Andaluc\'ia, CSIC, Glorieta de la Astronom\'ia s/n, E-18008 Granada, Spain}

\author[0000-0002-6907-0926]{Kei Tanaka}
\affiliation{Department of Earth and Planetary Sciences, Institute of Science Tokyo, Meguro, Tokyo, 152-8551, Japan
}

\author[0000-0001-6465-9590]{Zoie Telkamp}
\affiliation{Department of Astronomy, University of Virginia, Charlottesville, VA, 22904, USA}

\author[0000-0001-7511-0034]{Yichen Zhang}
\affiliation{Department of Astronomy, Shanghai Jiao Tong University, 800 Dongchuan Rd., Minhang, Shanghai 200240, People’s Republic of China}

\author[0000-0002-7299-8661]{Christian Fischer}
\affiliation{IRAM - Institut de Radioastronomie Millimétrique, 300 rue de la Piscine, 38406 Saint-Martin d'Hères, France}

\author[0000-0002-2986-2304]{Lianis Reyes Rosa}
\affiliation{Department of Astronomy, University of Virginia, Charlottesville, VA, 22904, USA}

\begin{abstract}
Massive stars regulate galaxy evolution and star formation through their physical and chemical feedback, but their formation remains poorly understood. Accretion-powered outflows provide important diagnostics of massive star formation. We present first results from the SOMA Atomic Outflow Survey, a far-infrared massive star formation survey using the FIFI-LS instrument on SOFIA. We report detection of \OIII\ $^3P_2\rightarrow^3P_1$ emission at 52 \micron\ from the massive protostar G11.94-0.62, tracing highly ionized gas. We also detect \OI\ $^3P_2\rightarrow^3P_1$ and $^3P_1\rightarrow^3P_0$ at 63 and 145 \micron\ tracing atomic gas, as well as CO $J=14\rightarrow13$ at 186 \micron\ from highly excited molecular gas. The \OIII\ and \OI\ lines exhibit large line widths ($\sim200$ and $\sim40-80$ \kms, respectively) and their morphologies are consistent with a wide-angle bipolar outflow. The properties of molecular tracers ($^{12}$CO, $^{13}$CO, C$^{18}$O, H$_2$CO, and CH$_3$OH) observed with ALMA support this interpretation. Ionized nebula and PDR modeling imply an ionized outflow mass flux of $\sim8\times10^{-5}\:M_\odot$ yr$^{-1}$ and an atomic outflow mass flux of $\sim5\times10^{-6}\:M_\odot$ yr$^{-1}$, while the molecular outflow traced by CO has an implied mass flux of $\sim3\times10^{-4}\:M_\odot$ yr$^{-1}$. The mass and momentum flux in the ionized outflow are consistent with the primary disk wind, while the molecular component is mainly swept-up, secondary outflow gas. We also observe G11.94-0.62 with the LBT in the near-infrared, potentially tracing the base of wide-angle outflow cavities. SED modeling implies a protostellar mass $m_* = 22.4^{+21}_{-11}\:M_\odot$, while the \OIII\ emission implies $m_*\gtrsim30\:M_\odot$ and that the protostar is in the final stages of its accretion.
\end{abstract}

\section{Introduction}

Massive stars impact many aspects of the evolution of galaxies, the interstellar medium, and star and planet formation \citep{2007arXiv0712.1109B, 2014prpl.conf..149T}. However, determining the formation processes of massive stars is often challenging due to the presence of complex, high column density gas and dust structures and the large distances to the sources. Bipolar protostellar outflows, ubiquitous in both low- and high-mass star-forming regions, provide an important window to constrain the properties of protostars \citep{2016ARA&A..54..491B}. The low-density, low-extinction outflow cavities can enable observations at shorter, near and mid infrared wavelengths, to probe into the inner regions of protostellar cores. In addition, the kinematics of these accretion powered outflows, e.g., their mass and momentum flux, can serve as important diagnostics for understanding protostellar accretion. Finally, understanding the outflows in their own right is important since they likely regulate the local star formation efficiency from the massive protostellar core and help regulate surrounding star formation activity in the protocluster clump.

One key difference of massive protostars compared to their lower-mass counterparts is the potential for much stronger far- and extreme-ultraviolet (FUV \&\ EUV) radiation. This phenomenon is expected to become important in the phase when massive protostars contract to near the zero-age main sequence while still accreting, which is theorized to occur for the most massive sources ($m_*\gtrsim 20\,M_\odot$) \citep{2014ApJ...788..166Z}. Irradiated by FUV and EUV photons, outflowing gas can be dissociated and ionized, producing strong lines that diagnose atomic and ionized phases \citep[e.g.,][]{2016ApJ...818...52T}. Thus, studying protostars exhibiting such ionized and atomic outflows is important for understanding massive star formation at the onset of powerful radiative feedback.

G011.94-00.62 (hereafter G11.94) is an ultra-compact \HII\ region whose morphology was first described by \citet{1989ApJS...69..831W}. We adopt a distance of 4.02~kpc and a $v_\text{lsr}$ of 37.0~km~s$^{-1}$ \citep{2015A&A...579A..91W}.  From VLA 2 and 6 cm images, \citet{1989ApJS...69..831W} identified a parabolic `cometary' ionization front that peaks toward the southwest side of the protostar. Early observations at millimeter wavelengths found compact dust emission without a clear structure \citep{1999ApJS..125..143W}.  Wide-field Midcourse Space Experiment (MSX) observations in the mid-IR (6--25 \micron) identified a slight elongation to the NW and SE and also detected a faint source 70\arcsec\ to the southwest \citep{2003MNRAS.343..143C}. A similar NW-SE elongation is seen in mid-IR IRTF/MERLIN observations at 11.7 and 20.8 \micron\ \citep{2003ApJ...598.1127D}, but the mid-IR and radio emission have different brightness peaks and the cometary shape seen in radio observations is missing in the mid-IR images.  \citet{2008ApJS..177..584Z} presented the distribution of [Ne\,\textsc{ii}] 12.8 \micron\ line, showing a broad line width with most [Ne\,\textsc{ii}] emission appearing between $v_\text{lsr}=30-60$ \kms.  However, a detailed kinematic analysis was challenged by the complex morphology of the system.

\citet{1996A&AS..120..283H} reported two regions of water maser emission, located $\sim3\arcsec$ SW and $\sim10\arcsec$ W from the UC $\HII$ region. They found both masers to be located to the west of the large cometary arc. Both water maser locations were later found to have corresponding mid-IR sources \citep{2003ApJ...598.1127D}. \citet{2003A&A...410..597W} associated a methanol maser with one of the water maser sites.

Regarding molecular emission, a single-pointing $^{12}$CO $J=1\rightarrow0$ observation by \citet{1996ApJ...457..267S} shows a broad full-width zero-intensity line width greater than 28.5 \kms, consistent with that seen in [Ne~\textsc{ii}], indicating the presence of an outflow.  Single-dish observations of $^{13}$CO $J=1\rightarrow0$ also found a broad line with three peaks, suggesting complex structure within the single-dish beam of 22\arcsec\ \citep{1992A&A...253..541C}. 

In this paper, we present observations of atomic and ionized outflows in G11.94, taken with the FIFI-LS instrument onboard SOFIA. In \S\ref{sec:obs}, we describe our FIFI-LS observations and reduction process, as well as discussing archival ALMA data and new LBT imaging which supplements our SOFIA observations. In \S\ref{sec:results}, we present the spectra for the \OI\ 63 \micron\ and 145 \micron\ lines, the \OIII\ 52 \micron\ line, and the CO $J=14-13$ line at 186 \micron. We then present the morphologies for each line as well as their associated continua. In \S\ref{sec:discussion}, we present a discussion on the outflow position angle, kinematics, and origins of the detected species, as well as the \OI\ mass loss (outflow) rate estimated using non-LTE radiative transfer calculation and PDR modeling. We also refer here to ALMA data for estimates of the molecular outflow. In \S\ref{sec:conclusion}, we discuss the implications of our results and present our conclusions.

\section{Observations}\label{sec:obs}

\subsection{SOFIA FIFI-LS}\label{sec:obs_fifils}

Observations were taken using the Far-Infrared, Field-Imaging Line Spectrometer \citep{2014SPIE.9147E..2XK, 2018JAI.....740003F, 2018JAI.....740004C} attached to the 2.5 meter mirror (aperture) onboard the Stratospheric Observatory for Infrared Astronomy (SOFIA) \citep{2018JAI.....740011T}. The program (project ID 09\_0169; PI: Y.-L. Yang) covers 18 massive protostars selected from the larger sample of the SOFIA Massive (SOMA) Star Formation Survey (PI: J. C. Tan) to survey the emission of the \OI\ $^3P_2\rightarrow^3P_1$ and $^3P_1\rightarrow^3P_0$ transitions at 63.18 $\micron$ and 145.5 $\micron$ \citep{1991ApJ...371L..85Z}, the \OIII\ $^3P_2\rightarrow^3P_1$ at 51.81 $\micron$ \citep{1985PhyS...32..181F} and the CO $J=14\rightarrow13$ line at 185.99 $\micron$ \citep{2010ApJ...718.1062Y}. For brevity, we will hereafter refer to these lines as \OI\ 63 and 145 \micron, \OIII\ 52 \micron, and CO 186 \micron, respectively. In this work, we report the SOFIA/FIFI-LS data on G11.94. The observations were taken on 2021 June 3, with exposure lengths (on source) of 460.8~s for the 63-145 \micron\ bandpair and 307.2~s for the 52-186 \micron\ bandpair.

\begin{figure*}[htbp!]
    \centering
    \includegraphics[width=\textwidth]{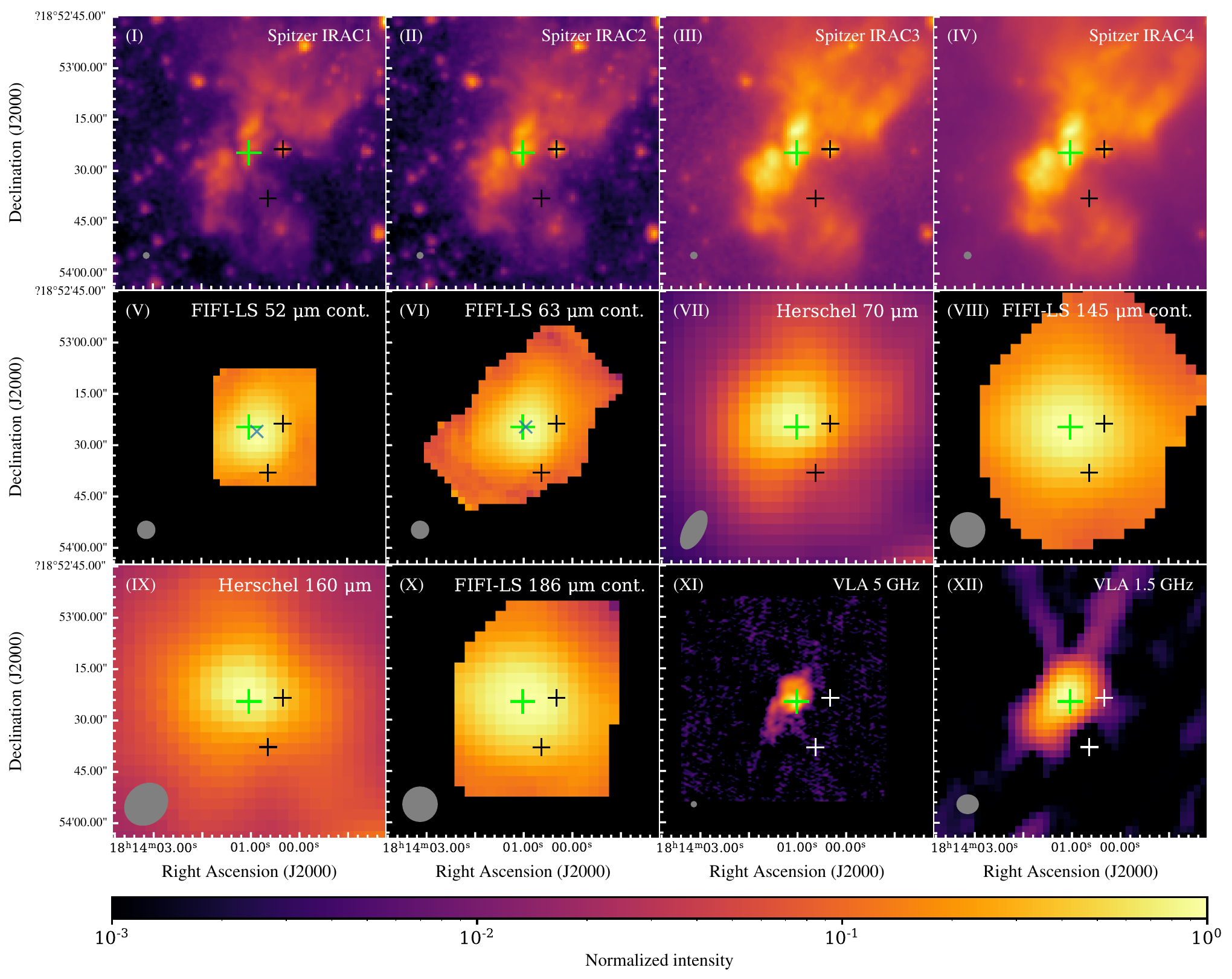}
    \caption{Continuum images of G11.94 from the various wavelengths observed by SOFIA FIFI-LS (this work), Spitzer \citep{2009PASP..121..213C}, Herschel (\citealt{2016A&A...591A.149M}, OBSIDs: 1342218999, 1342219000), and VLA \citep{1994ASPC...61..165B, 2012PASP..124..939H}.  The images are shown in logarithmic scale, normalized to the highest flux in an $80\arcsec\times80\arcsec$ region centered on the source peak. The green cross denotes the source identified as the G11.94 protostar from ALMA continuum. The two black crosses (white in panels XI and XII for clarity) are secondary protostars in the region (both detected with ALMA; see \S\ref{sec:origin}). Panels V and VI also show the peak of emission at 52 and 63 \micron\ (blue `$\times$' symbols). The offset between green `+' and blue `$\times$' helps define the ``color gradient'' (see text). Beam sizes are shown in the bottom left of each panel.}
    \label{fig:multipanel}
\end{figure*}

FIFI-LS is an integral field unit comprised of two spectrometers, capable of observing the same region simultaneously at two wavelength channels using a dichroic. The short-wavelength (hereafter SW) channel covers wavelengths from 51--120 $\micron$ and the long-wavelength (hereafter LW) channel covers 115--200 $\micron$.  Both detectors in the SW and LW channels have 5$\times$5 spatial pixels with a size of 6\arcsec\ and 12\arcsec\ respectively, and 16 spectral pixels each. Thus, the SW and LW channels have an instantaneous field-of-view (FoV) of $30\arcsec\times30\arcsec$ and $60\arcsec\times60\arcsec$ with beam sizes of 6.6, 7.4, 14.1, and 18.1\arcsec\ in the 52, 63, 145, and 186 \micron\ ranges.  We set up two observations in which each band-pair (63+145 \micron\ and 52+186 \micron) was simultaneously observed.  In our observations, we use the D130 and D105 dichroics to maximize the throughput at 52 and 63 \micron, respectively. The velocity resolution of FIFI-LS is 320, 230, 280, and 200 \kms\ in the 52, 63, 145, and 186 \micron\ ranges, respectively \citep{Fadda_2023}. The instantaneous spectral coverage is 1000–3000 \kms\ in SW channels and 800–2500 \kms\ in LW channels.

The data were processed with version 2.9.0 of the FIFI-LS pipeline provided by the SOFIA Science Center and the atmospheric correction with ATRAN \citep{Lord92} outlined by \citet{2021PASP..133e5001F} and \citet{2021PASP..133e5002I}. The pipeline uses a 3-D fitting method to bin the data, which rejects data points from bad fits. This affects the edge shape of the reduced maps due to lower integration times there caused by dithering \citep{fischer16} during the mapping (see panels VI and VIII in Fig. \ref{fig:multipanel}). There are known telluric features in all of the channels (except for 186 \micron), and our methods for handling these features are explained throughout the paper. In the 52 \micron\ channel (panel V), this manifests as a large peak to the blue-side of the line, beginning roughly around $-400\:$\kms\ and beyond. This region is excluded from line measurements, and we manually avoid selecting any wavelength from it for measuring the continuum. The 145 \micron\ band (panel VIII) features a small `bump' around $-500\:$\kms\ and similarly this region of the spectrum is excluded from inclusion in any analysis, including continuum selection. The situation for the 63 \micron\ spectrum (panel VI) is slightly different. The telluric feature convolves directly with the line on the red wing of the spectrum. Our methods for resolving this contamination are outlined in Section\,\ref{sec:line_emission}.

\subsection{ALMA}

We also obtained archival ALMA (Atacama Large Millimeter/submillimeter Array) data to investigate the nature of the emission lines detected in our FIFI-LS observations (discussed in \S\ref{sec:origin}).  The $^{12}$CO data were taken as part of the QUARKS program (2021.1.00095.S; PIs: L. Zhu, G. Gary, and T. Liu; \citealt{2024RAA....24b5009L}), using the 12m-array in the C-2 and C-5 configurations, as well as the ACA (Atacama Compact Array) 7m-array \citep{2024RAA....24f5011X}.  The calibrated visibilities of these data were combined using CASA 6.6.0 \citep{2022PASP..134k4501C} and deconvolved together.  The details of the reduction and imaging processes are discussed in \citet{2024RAA....24b5009L}. The synthesized beam is $\sim$0\farcs{3} with a sensitivity of $\sim$6~mJy~beam$^{-1}$. The other ALMA data were taken as part of the ALMAGAL: ALMA Evolutionary study of High Mass Protocluster Formation in the Galaxy program (2019.1.00195.L; PI: S. Molinari; \citealt{2025A&A...696A.149M}) and the imaging products were gathered from the ALMA Science Archive without further calibration and imaging. Four spectral windows were set up in the ALMA observations of G11.94, covering frequencies from 216.960--218.832~GHz, 218.037--218.505~GHz, 219.032--220.905~GHz, and 220.337--220.805~GHz with a channel width of 488~kHz, 122~kHz, 488~kHz, and 122~kHz, respectively.  The synthesized beam sizes are (1\farcs{31}--1\farcs{37})$\times$(1\farcs{06}--1\farcs{08}).  The sensitivity is $\sim$6 mJy beam$^{-1}$.

\subsection{LBT}

Observations were conducted on July 24 and 26, 2025, using the Large Binocular Telescope (LBT) in binocular mode, which allows for simultaneous use of both mirrors. The data were collected with the LUCI (LBT Utility Camera in the Infrared, \citet{seifert03}) instrument operating in seeing-limited conditions as part of program UV-2025A-03 (PI: J. C. Tan). The N3.75 camera was employed, offering a field of view of $4\arcmin \times 4\arcmin$ and a pixel scale of 0.12\arcsec. On LUCI1, installed on the SX (left) mirror, observations were made using the K and Br$\gamma$ filters (centered at 2.194 and 2.170~$\mu$m, respectively). LUCI2, mounted on the DX (right) mirror, used the K and H$_2$ filters (centered at 2.194 and 2.124~$\mu$m, respectively). The images are centered at RA(J2000)=18:14:01.100 and Dec(J2000)=$-$18:53:23.53, with a position angle of 270$^\circ$. Exposure times were 300 seconds for the K-band and 600 seconds for the Br$\gamma$ and H$_2$ narrow-band filters. Data reduction followed standard procedures, including sky subtraction and flat-fielding, using custom Python routines based on the ccdproc \citep{ccdproc}, astropy \citep{astropy} and photutils \citep{photutils} packages. Final images were aligned and calibrated astrometrically by cross-matching stellar positions with the 2MASS catalog \citep{2006AJ....131.1163S}.

\section{Results}\label{sec:results}

Using SOFIA/FIFI-LS, we obtained a spectral cube at each line. Massive protostellar cores such as G11.94 emit strong continuum emission at far-infrared wavelengths \citep[e.g.,][]{2017ApJ...843...33D, 2019ApJ...874...16L, 2020ApJ...904...75L, 2023ApJ...942....7F, 2025ApJ...986...15T}. To isolate the emission lines, we need to first measure the continuum at each channel and subtract it from the spectral cubes. In this section, we present the results on continuum imaging/flux measurements in \S\ref{sec:continuum} and various line emission science in \S\ref{sec:line_emission}.

\subsection{Continuum}\label{sec:continuum}

We measure the continuum from the line-free wavelengths in the bandwidth of each channel. Thus, we select continuum ranges from outside of 300~\kms\ of the source $v_\text{lsr}$ surrounding the line. The exclusion zone is to ensure there is no contamination from line emission in the continuum, particularly due to the relatively limited spectral resolution of FIFI-LS. We also only select continuum from where atmospheric interference is at a minimum. This means excluding continuum selection from the blue side of the spectra in \OIII\ and the red side in \OI\ 63 \micron\ due to telluric features, and partially excluding regions from the blue side of the spectra in \OI\ 145 \micron. Specifically, the continuum was selected from the following ranges (relative to line center; all in \kms): 52 \micron: 450 to 1050; 63 \micron: -650 to -300; 145 \micron: -450 to -350 and 350 to 1000; 186 \micron: -550 to -300 and 350 to 450. Then, for each pixel, we measure the average flux density within the selected wavelengths for continuum, deriving a continuum image. These continuum images are later subtracted from the corresponding spectral cubes at each wavelength to produce line-only spectral cubes for line analysis.

\subsubsection{Astrometric Correction}

We further compare the FIFI-LS continuum images at 52, 63, 145, and 186~\micron\ with archival data on G11.94, including images from the Spitzer Space Telescope (Spitzer), the Herschel Space Observatory (Herschel), the Very Large Array (VLA), and ALMA. To provide an accurate comparison, we need to ensure the astrometry of the FIFI-LS images is consistent with other images. To do so, we first fit a 2-dimensional Gaussian profile onto the ALMA continuum image at 218.935~GHz, determining the continuum peak position at (R.A.: 18$^h$14$^m$01\fs{00}, Decl.: $-$18\arcdeg53\arcmin24\farcs{96}). We also use this ALMA continuum peak as the source position in the rest of the analyses.  Then, we fit a 2-D Gaussian profile to each FIFI-LS continuum image to determine the peak positions. To align the images, we only offset the fitted FIFI-LS continuum peak at the two LW channels (145 and 186 \micron) to match the position of the ALMA continuum peak.  Because each pair of SW and LW channels was observed simultaneously, we applied the offsets of the LW channels to the corresponding SW channels, respectively. The rationale for correcting the LW channel instead of the SW is that the dust is more optically thin at longer wavelengths, and the continuum at shorter wavelengths could be affected by selective extinction due to asymmetric source structures.

The simultaneous observations of the SW and LW channels also provide a unique opportunity to study the effect of this selective extinction.  If there is a dominant bipolar outflow, we expect a lower extinction at the blue-shifted outflow lobe and a higher extinction at the red-shifted outflow lobe, resulting in the continuum peak shifting toward the blue-shifted outflow at shorter wavelengths \citep{2011ApJ...733...55Z}.  We derive the ``relative offset'' between the continuum peak at the SW and LW channel of the same observing pair to quantify this selective extinction effect, the position angle (PA$_{\text{cont}}$) of which could also serve as an indicator of outflow direction (see further discussion in \S\ref{sec:orientation_FIFI}).

Figure\,\ref{fig:multipanel} shows the FIFI-LS continuum images along with archival photometric images from 3.6 \micron\ to 20 cm.  The relative offsets are shown as blue crosses in panels V and VI. The 63-145 \micron\ relative offset and P.A. are $1.0\pm0.3\arcsec$ and $-88.5\pm19.3\degree$ and the 52-186 \micron\ relative offset and P.A. are $2.5\pm0.5\arcsec$ and $-126.2\pm12.4\degree$.

\subsubsection{Continuum Measurements}

The spectral energy distribution (SED) of massive protostellar sources typically peaks at $\sim100$ \micron\ because of the abundant relatively cool dust ($\sim20-100$ K) in the envelopes.  Thus, the continuum fluxes at 50--200 \micron, probed by the FIFI-LS observations, also constrain the SED of G11.94 and further constrain the properties of the dominant protostar by SED modelling (\S\ref{sec:seds}).  To estimate the continuum flux for modelling, we first need to determine an aperture that best encompasses the emission of the entire protostellar core.  We followed the approach developed in \citet{2023ApJ...942....7F}, which uses the Herschel 70 \micron\ image to determine the aperture where a 30\%\ increase of aperture radius leads to less than 10\%\ increase of flux. Following this criterion, the optimal aperture radius is 30.0\arcsec.

\begin{table}[]
    \centering
    \begin{tabular}{c|c|c|c}
        \hline
        \hline
        \OIII\ 52 $\micron$ &  \OI\ 63 $\micron$ &  \OI\ 145 $\micron$ &  CO  186 $\micron$ \\
        \hline
        $1244.8 \pm 53$ & $1494.0 \pm 42$ & $1110.1 \pm 24$ & $1054.6 \pm 29$ \\

        \hline 
        \hline  
    \end{tabular}
    \caption{Background-subtracted continuum fluxes (in Jy) for each bandpass observed with FIFI-LS.}
    \label{tab:continuum_fluxes}
\end{table}

\begin{figure*}[htbp!]
    \centering
    \includegraphics[width=\textwidth]{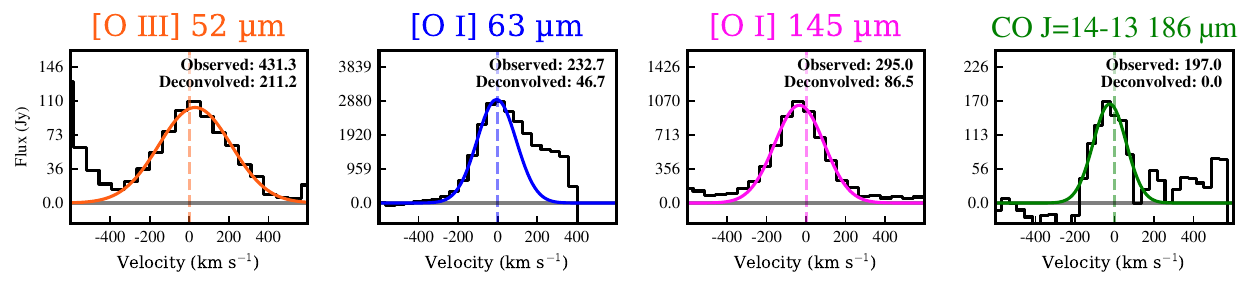}
    \caption{Continuum-subtracted spectra of the \OIII\ $^3P_2\rightarrow^3P_1$ at 51.81 $\micron$, \OI\ $^3P_2\rightarrow^3P_1$ and $^3P_1\rightarrow^3P_0$ at 63.18 $\micron$ and 145.5~$\micron$, and the CO $J=14\rightarrow13$ line at 185.99~$\micron$. The velocity is corrected for the source $v_{\rm lsr}$ of 36.2~\kms. The fitted Gaussian lines are shown. For \OI\ in the 63 \micron\ bandpass, the Gaussian profile is fitted over a partial range due to telluric contamination on the red wing. Observed line FWHMs (in \kms) are written in the top right of each panel, as well as FWHMs deconvolved from the intrinsic instrument FWHM. The intrinsic width of the CO line is not able to be ascertained via deconvolution and is listed as zero. Flux tickmarks are plotted in 33\%\ increments of the peak line flux.}
    \label{fig:spectra}
\end{figure*}

The measurement of continuum flux also follows the approach as that in \citet{2023ApJ...942....7F}.  A background-subtracted flux is measured from the 60\arcsec\ aperture using \texttt{SedFluxer} from the Python package \texttt{sedcreator}, while the background is measured from an annulus between one and two aperture radii ($r=30\arcsec-60\arcsec$). We applied this approach to the continuum images at Spitzer IRAC bands (3.6, 4.5, 5.8, and 8.0~\micron), WISE Band 3 and 4 at 12 and 22~\micron, four wavelengths of the FIFI-LS observations presented in this study, Herschel PACS 70 and 160~\micron\ (which can be seen in panels VII and IX in Fig.~\ref{fig:multipanel}), and Herschel SPIRE 250, 350, and 500~\micron. In the 145 and 186~\micron\ channels, the aperture size is comparable to the FoV of the LW channel, which is only a few pixels smaller after masking out the bad pixels.  However, the FoV at 52 and 63~\micron\ is substantially smaller than the aperture. Thus, we interpolate the area outside the FoV of the SW channel using the Herschel 70 \micron\ to estimate the continuum flux within the aperture.  This is accomplished by reprojecting the Herschel 70 \micron\ image to the FIFI-LS pixel size/orientation, and then calculating the proportion of fluxes within the FIFI-LS FOV in both images. The Herschel 70~\micron\ image is then scaled using this factor, where FIFI-LS data is not present. It is seen from our larger survey sample that this slightly overestimates flux at these two wavelengths and, as such, the 52 and 63~\micron\ fluxes are treated as upper limits in our SED fits. We similarly interpolate data from Herschel 160~\micron\ to FIFI-LS 145 and 186~\micron\ measurements. Due to the larger FOV of these LW images, this interpolation only is relevant for small fractions of the annulus.

We estimate the uncertainties of all the continuum fluxes following the treatment described in \citet{2023ApJ...942....7F}. Specifically, for background subtracted fluxes of the protostar that are use in SED fitting we need to estimate the uncertainty in the background flux. At wavelengths $\geq100$ \micron\ the uncertainty in the background flux is assumed to be equal to the level of background flux itself, reflecting the fact that the background is relatively cold, clump material that can dominate over the protostellar core emission. However, at wavelengths $<100$ \micron\, the uncertainty in the background flux is assumed to be given by the level of variation seen in aperture-sized regions immediately around the source, which is derived using \texttt{SedFluxer}. The uncertainty in background flux is then added in quadrature with an assumed 10\%\ of the background-subtracted flux \citep{2018JAI.....740003F} to derive the total uncertainty. The SED and the properties inferred from the SED modeling are discussed in \S\ref{sec:seds}.

\subsection{Line Emission}\label{sec:line_emission}

\begin{table*}[htbp!]
    \begin{center}
    \centering
    \begin{tabular*}{0.61\textwidth}{c|c|c|c}
        \hline
        \hline
        \OIII\ 52 $\micron$ &  \OI\ 63 $\micron$ &  \OI\ 145 $\micron$ &  CO  186 $\micron$ \\
        \hline
        $7.72\pm0.45\times10^4$ &  $7.27\pm1.4\times10^5$ & $3.24\pm0.11\times10^5$ & $3.55\pm0.59\times10^4$ \\
        \hline 
        \hline  
    \end{tabular*}
    \caption{Line fluxes (in Jy km s$^{-1}$) obtained via 1-D Gaussian fitting of the FIFI-LS spectra.}
    \label{tab:line_fluxes}
    \end{center}
\end{table*}

We extract a spectrum for each channel from continuum-subtracted spectral cubes within a 60\arcsec\ aperture centered on the source  (Fig.~\ref{fig:spectra}).  The extraction is performed by using \texttt{aperture\_photometry()} from \texttt{photutils} at each wavelength. The continuum subtraction uses the image derived from the method described in \S\ref{sec:continuum} and deduces the continuum from each wavelength of the corresponding spectral cube.  Given the size of the aperture, the extracted spectra at 52 and 63 \micron\ are effectively the spectra of the entire FoV. 

\begin{figure*}[htbp!]
    \centering
    \includegraphics[width=\textwidth]{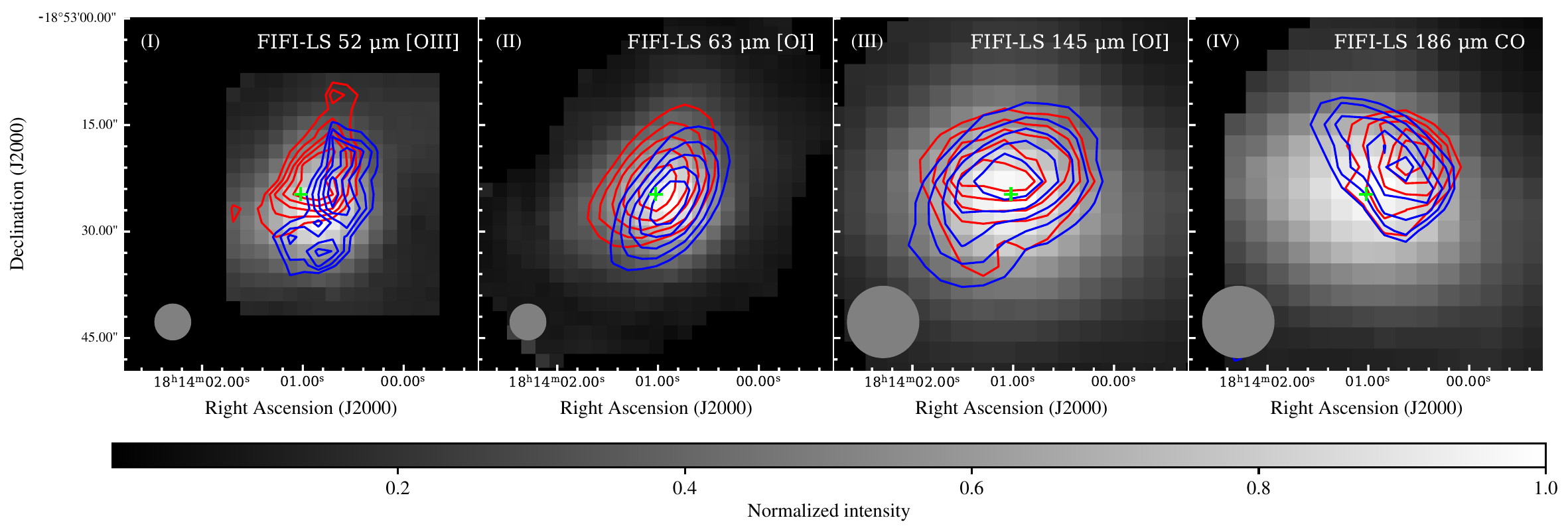}
    \caption{Blue- and red-shifted emission of the \OIII\ $^3P_2\rightarrow^3P_1$ at 51.81 $\micron$, \OI\ $^3P_2\rightarrow^3P_1$ and $^3P_1\rightarrow^3P_0$ at 63.18 $\micron$ and 145.5 $\micron$, and the CO $J=14\rightarrow13$ line at 185.99 $\micron$.  The velocity ranges are $\pm(50-300)$ \kms\ for every contour map shown.  Contours are shown from 50\%\ to 100\%\  with an increment of 10\%\ of the integrated flux for both blue- and red-shifted velocities, independently. The background images show the normalized continuum at corresponding wavelengths. The beam size is shown at the bottom left corner of each panel.}
    \label{fig:line_emission}
\end{figure*}

We measured the line fluxes by fitting a 1-D Gaussian profile to the extracted spectra, within $\pm300$~\kms\ of the source velocity. Due to the telluric features mentioned in \S\ref{sec:continuum}, we took special care on the line fitting, especially for the \OI\ 63~\micron\ line, where the red wing of the line blends with a prominent telluric feature.  To minimize the telluric contamination, we only fit the spectrum from $-$300 to 50~\kms.  The telluric feature at $\lesssim-500$~\kms\ is sufficiently separated from the \OIII\ 52~\micron\ line so that no additional treatment is required for the line fitting.

The fitted line fluxes are listed in Table~\ref{tab:line_fluxes}. The fitted line widths and the intrinsic line widths after deconvolving from the instrumental spectral resolution are shown in Figure~\ref{fig:spectra}. We note that the \OI\ 63~\micron\ line profile is likely to have narrow self-absorption, leading to an underestimation of its line flux when the absorption is not resolved \citep{2014AA...562A..45K,2015AA...584A..70L}.

Figure\,\ref{fig:line_emission} shows the distribution of line emission split into blue- and red-shifted velocities compared to the continuum at the same wavelength.  To investigate the kinematics of \OIII, \OI, and CO, we chose the velocity ranges to avoid telluric features as well as a central exclusion zone of $\pm\zeta$ \kms, where we define $\zeta$ to be equivalent to one sixth of the velocity resolution of FIFI-LS at the given bandpass. This ensures that any distribution of blue- or red-shifted emission is highlighted clearly and is not blurred by instrumental resolution. The blue- and red-shifted emission is clearly separated for the \OIII\ 52~\micron\ (despite the poor spectral resolution in this bandpass), while the separation for the \OI\ 63 \micron\ is notable but less significant.  The \OI\ 145 \micron\ and CO 186 \micron\ lines show no obvious morphological differences between the blue- and red-shifted emission.

\section{Analysis}\label{sec:discussion}

\subsection{Protostellar Properties from SED Fitting}\label{sec:seds}

Here we discuss properties of the protostar that can be inferred from its SED. In a series of papers \citep{2011ApJ...733...55Z,2013ApJ...766...86Z,2014ApJ...788..166Z,2018ApJ...853...18Z} a self-consistent model grid of massive protostellar cores has been developed, combining the evolution of the protostar and outflow. This model grid, based on the turbulent core accretion (TCA) model \citep{2002Natur.416...59M,2003ApJ...585..850M}, describes the protostellar structures using five parameters, $m_*$, $M_{c}$, $\Sigma_{\rm cl}$, $\theta_{\rm view}$, and $A_V$, which are the current protostellar mass, the initial core mass, the mean mass surface density of the surrounding clump, the angle of the line of sight to the outflow axis, and the foreground visual extinction.  This model grid allows us to constrain the protostellar properties by fitting the SED of G11.94, using the fitting tool \texttt{SEDFitter} within the Python package \texttt{sedcreator} \citep{2023ApJ...942....7F}.

\begin{figure*}[htbp!]
    \centering
    \includegraphics[height=2in]{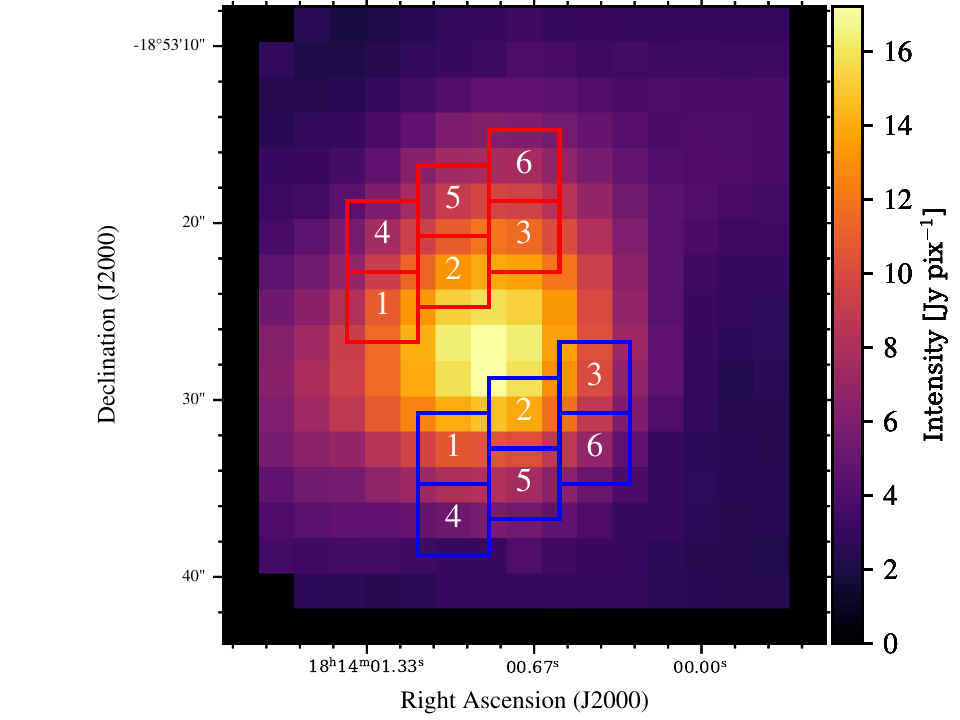}
    \includegraphics[height=2in]{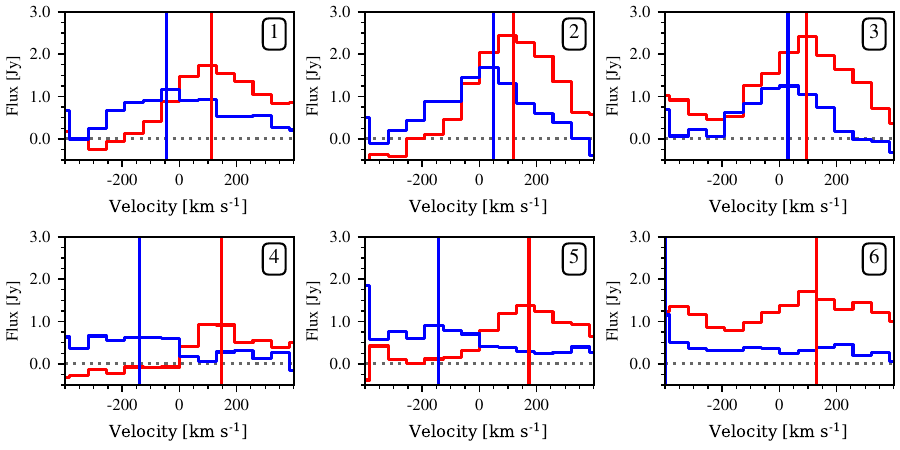}
    \caption{{\it Left:} Continuum image at 52 \micron\ highlighted with two sets of 2 sq. pixel apertures in the regions where blue- and red-shifted \OIII\ emission is detected.  {\it Right:} The continuum-subtracted spectra extracted from the apertures in the left panel.  Each panel shows the spectra extracted from a pair of apertures indicated by the number. These pairs of apertures are chosen to be at opposite sides from the source, probing the blue- and red-shifted lobes. The peaks of Gaussian profiles fitted from the red and blue wings of the spectra are shown as vertical color-coded bars to highlight the peak velocity of the spectra at both sides of the source, as well as the velocity offset between the two lobes.}
    \label{fig:kinematics}
\end{figure*}

We constructed the SED using the extracted continuum fluxes that include both FIFI-LS data and archival measurements (see \S\ref{sec:continuum}).  We follow the same approach in \citet{2023ApJ...942....7F} to treat the continuum fluxes at $\leq8 \micron$ as upper limits, because the model grid does not include polycyclic aromatic hydrocarbon (PAH) features, nor single photon transient heating effects on small dust grains \citep{2017ApJ...843...33D,2018ApJ...853...18Z}. Both processes could increase the fluxes at $\leq 8 \micron$. As mentioned in \S\ref{sec:continuum}, FIFI-LS continuum measurements at 52 and 63~\micron\ are treated as upper limits due to the possible overestimation of the flux from the interpolation using the Herschel 70~\micron\ image.

\begin{figure}[htbp!]
    \centering
    \includegraphics[width=0.5\textwidth]{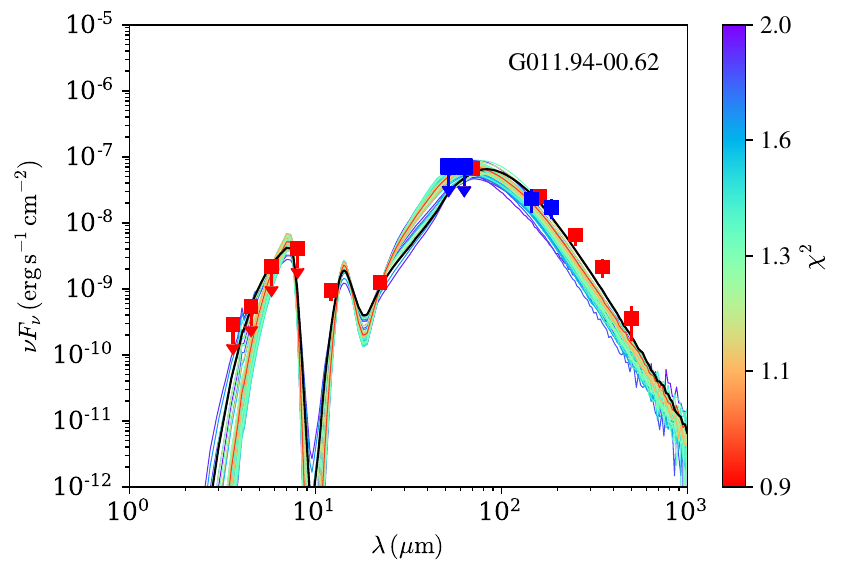}
    \caption{Spectral energy distribution for G11.94. The blue data points are measured by FIFI-LS, while the red points are from other observatories. The black line represents the best-fit model, with a $\chi^2$ value of 0.85.}
    \label{fig:sed}
\end{figure}

Utilizing criteria from \citet{2023ApJ...942....7F} for establishing which SED fit models are ``good'', we select the 44 models with the lowest $\chi^2$ values, ranging from 0.85 to 1.97, i.e., only models with $\chi^2$ less than 2 are considered. In addition, for physical self-consistency, models are required to have an initial core radius, $R_c$, less than twice the aperture radius \citep[see discussion by][]{2023ApJ...942....7F}. This is to ensure that the modeled radius does not exceed the extent of emission as seen in the 37 \micron\ FORCAST imaging used to derive the aperture. For explicit details on which models are considered ``good'', please refer to \citet{2023ApJ...942....7F}. We average across all these ``good'' models to retrieve estimates for protostellar properties (Table \ref{tab:SED_results}).

\begin{table*}[]
    \centering
    \begin{tabular*}{0.85\textwidth}{c|c|c|c|c|c|c|c}
        \hline
        \hline
        $m_\star$ [$M_\odot$] & $M_{c}$ [$M_\odot$] & $\dot{m}_\star$ $[M_\odot$ yr$^{-1}]$  & $L_{\rm bol}$ [$L_\odot$] & $R_{c}$ [pc] & $A_V$ [mag] & $\theta_{\rm view}$ [\degree] & $\Sigma_{\rm cl}$ [g cm$^{-2}$] \\
        \hline         
        $22.4^{+21}_{-11}$ &
        $223.1^{+92}_{-65}$ &
        $9.1^{+5}_{-3}\times10^{-4}$ & 
        $1.6^{+3}_{-1}\times10^5$ &$0.083^{+0.04}_{-0.02}$ &
        $287.7\pm{71.5}$ & 
        $17.3\pm{8.8}$ & 
        $1.74^{+1.4}_{-0.77}$ \\
        \hline
        \hline

    \end{tabular*}
    \caption{Best-fitting values of protostellar parameters obtained via SED fitting.}
    \label{tab:SED_results}
\end{table*}

\subsection{Origins of Ionized and Atomic Emission}
\label{sec:origin}

The detection of the \OIII\ $^3P_2\rightarrow$$^3P_1$ line (E$_\text{u}=440.5$~K) implies the presence of highly ionized  gas. The ionization of \OIII\ requires an energy of 35.12~eV \citep[i.e., from singly ionized oxygen,][]{Ingvar_Wenåker_1990,10.1063/1.555928}. If due to photoionization, this requires a stellar spectrum that is able to fully ionize He to He$^+$, i.e., achieved once photospheric temperatures reach about 37,000~K, which for zero age main sequence (ZAMS) stars is equivalent to approximately an O7 star with a mass of about $30\:M_\odot$ \citep{2005A&A...436.1049M,2011piim.book.....D}. In the context of the TCA model, a protostar of this mass is expected to reach the ZAMS as long as the mass surface density of the clump environment is $\Sigma_{\rm cl}\lesssim3\:{\rm g\:cm}^{-2}$ \citep{2014ApJ...788..166Z}. It is interesting to note that these requirements of $m_*\gtrsim30\:M_\odot$, and $\Sigma_{\rm cl}\lesssim3\:{\rm g\:cm}^{-2}$ are approximately consistent with those derived from SED modeling, described above.

Nevertheless, \OIII\ could also be produced via shock ionization. However, this is achieved for a fairly narrow range of temperatures at $\sim 9\times10^4\:$K \citep[e.g.,][]{1998A&AS..133..403M,2011piim.book.....D}. Below temperatures of $\sim6\times 10^4\:$K, most oxygen is in ionization state \mbox{[O~\textsc{ii}]}, while above temperatures of about $1.3\times10^5\:$K, most oxygen is in ionization state \mbox{[O~\textsc{iv}]}. Note, post shock temperatures of 0.6, 0.9 and $1.3\times10^5\:$K are achieved for shock speeds of 64, 79 and 95~\kms, respectively (assuming strong shock jump conditions for mean particle mass $0.636\:m_p$, i.e., for singly ionized He, and ignoring effects of magnetic fields).

As discussed in previous sections, the emission of \OIII\ is kinematically resolved by our FIFI-LS observations (Figure\,\ref{fig:spectra}).  The intrinsic (deconvolved) FWHM of the \OIII\ line is about 200~\kms, but with maximum velocities extending to $\pm400\:{\rm km\:s^{-1}}$. We note that the escape speed from a $30\:M_\odot$ ZAMS star, i.e., with radius of about $8\:R_\odot$, is about $1,200\:$\kms, which is representative of the maximum terminal velocity of an X-wind \citep{2000prpl.conf..789S} or inner disk wind \citep{2000prpl.conf..759K}. Especially for disk wind models, in which material is launched from a range of radii, there is expected to be a broad range of outflowing gas velocities up to this maximum. If we adopt an observed maximum 3D \OIII\ outflow velocity limit of $400\:{\rm km\:s}^{-1}$, i.e., assuming a negligible reduction due to inclination effects and deceleration due to sweeping up of ambient gas, then the fact this is about 1/3 of maximum outflow speed expected from a $30\:M_\odot$ ZAMS star may indicate that the innermost region of \OIII\ outflow launching occurs at $\sim10$ stellar radii. This may imply that the accretion disk is truncated at this location, e.g., by a strong stellar magnetic field, or that the fastest component of the outflow is not well traced by \OIII\ emission. We also note that the \OI\ emission lines appear to be narrower than the \OIII\ line, i.e., with FWHM of 46.7 \kms for the 63 \micron\ line and 86.5~\kms for the 145 \micron\ line. Their velocities appear to extend to about $\pm250\:$\kms. This indicates the atomic component of the outflow is slower than the ionized one, which is expected if it is launched from larger disk radii and/or suffering from a greater degree of deceleration. Thus, overall, the observed line of sight kinematics of \OIII\ and \OI\ appear consistent with the velocity profile expected of a structured ionized-atomic disk wind from a $\sim20-30\:M_\odot$ protostar that is near the ZAMS, although it is possible that the very fastest components of are not well traced by our observations.

Examining the morphology of the blue- and red-shifted \OIII\ emission, we see that this shows a clear velocity gradient in the northeast-southwest direction, which could also indicate a bipolar outflow (Fig. \ref{fig:line_emission}). In Figure\,\ref{fig:kinematics}, we investigate the peak of the \OIII\ line emission in the blue- and red-shifted areas highlighted by the contours in Figure\,\ref{fig:line_emission}. Each panel in Figure\,\ref{fig:kinematics} (right) shows the \OIII\ spectrum at two opposing positions from the source, roughly oriented along the outflow axis. The emission from the red-shifted region consistently peaks at a red-shifted velocity, and that from the blue-shifted region at a blue-shifted velocity.
We discuss the implications for this velocity gradient in \S\ref{sec:orientation_FIFI}.

In addition, the velocity distribution in the \OI\ $^3P_2\rightarrow^3P_1$ line at 63 \micron, despite a less prominent separation between the blue- and red-shifted velocities, corroborates the velocity gradient seen in the \OIII\ line.  We also expect the morphology of the \OI\ 145 \micron\ line (E$_\text{u}=326.6$ K) to be similar to that of the \OIII\ line.  However, the spatial resolution at 145 \micron\ is more than twice as poor as that at 52 and 63 \micron, which hinders a confirmation of bipolar outflowing gas using the \OI\ 145 \micron\ line. Finally, we note that the CO transition mapped by FIFI-LS appears to trace the innermost region of the protostellar environment, but, in addition to having a line width that is consistent with instrumental broadening, shows no hint of spatial velocity gradient. The CO emission may therefore trace either a much slower component of the outflow, or gas in the inner infall envelope or accretion disk.

\subsection{Additional Outflow Tracers}

\begin{figure}
    \centering
    \includegraphics[width=0.47\textwidth]{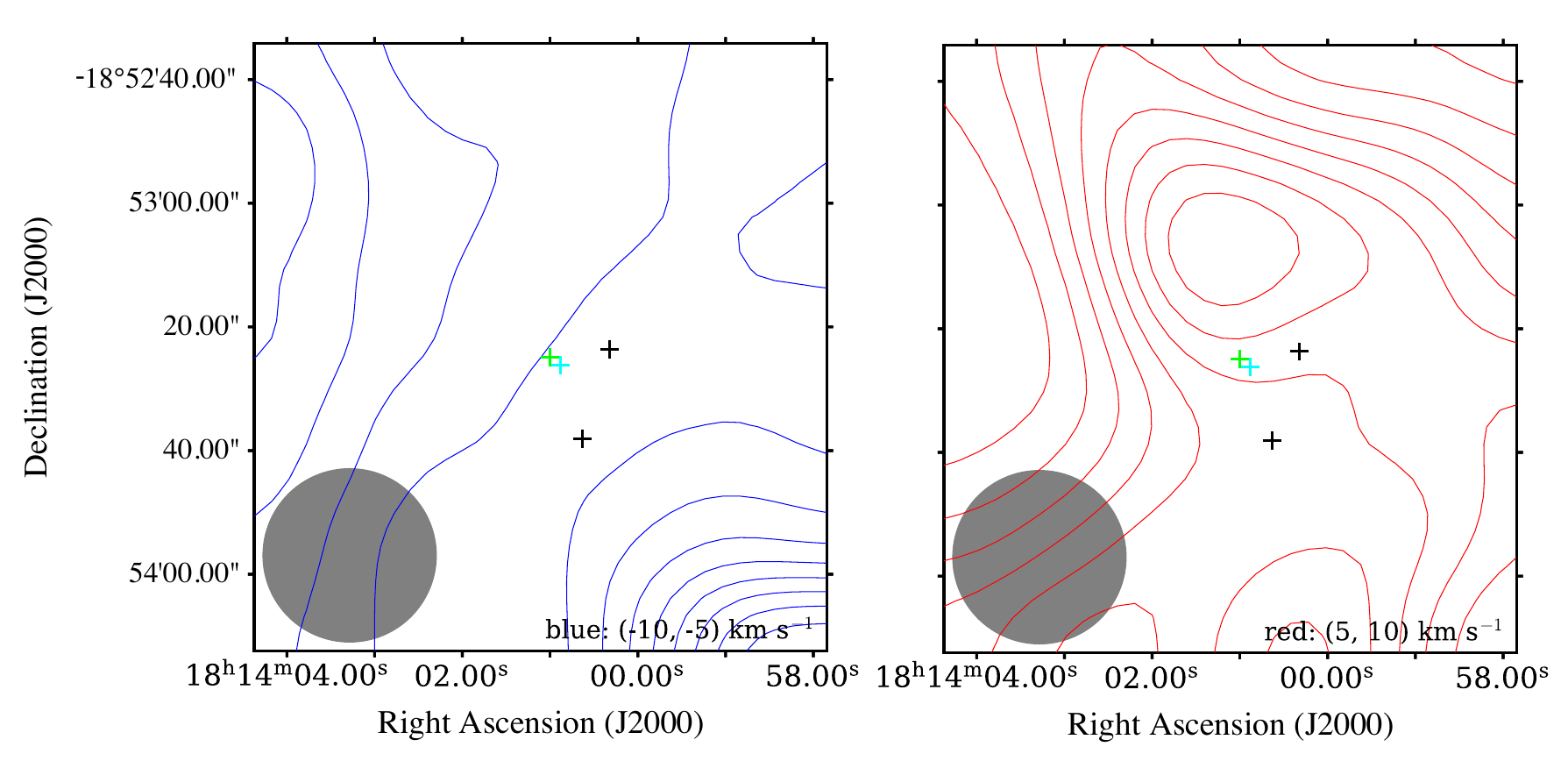}
    \caption{$^{12}$CO mapped with the ALMA total power antennas. Note the large beam size (bottom left). The peaks of red-shifted (NE) and blue-shifted (SW) emission trace the suggested $-145\degree$ outflow axis.}
    \label{fig:tp}
\end{figure}

We have found that the \OIII\ and \OI\ emission observed by SOFIA/FIFI-LS has a plausible origin via photoionization and photodissociation of a disk wind outflow from a massive protostar. Here we consider other evidence for a protostellar outflow from G11.94.

First, we examine archival ALMA data of G11.94, focusing on the $^{12}$CO $J=2\rightarrow1$, $^{13}$CO $J=2\rightarrow1$, C$^{18}$O $J=2\rightarrow1$, H$_2$CO $3_{03}\rightarrow2_{02}$, and CH$_3$OH $4_{22}\rightarrow3_{12}$ species and transitions, searching for signatures of outflows. First, examining ALMA total power (TP) observations of $^{12}$CO(2-1) (Fig. \ref{fig:tp}), we see evidence for redshifted emission (i.e., velocities $>5\:$\kms to the red of $v_{\rm lsr}$) that peaks to the NE of the source, while equivalently blueshifted emission is to the SW. This axis is consistent with that implied by the \OIII\ emission, discussed above.

\begin{figure*}
    \centering
    \includegraphics[width=\textwidth]{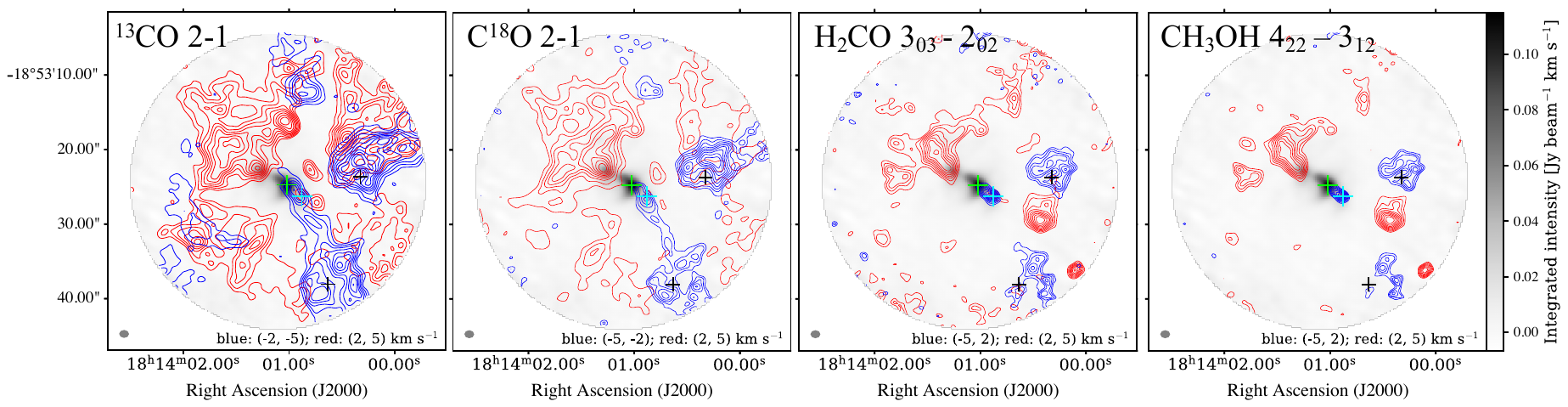}
    \caption{Maps of the $^{13}$CO $J=2\rightarrow1$,  C$^{18}$O $J=2\rightarrow1$, H$_2$CO 3$_{03}\rightarrow2_{02}$, and CH$_3$OH $4_{22}\rightarrow3_{12}$ lines as observed by ALMA.  The contours show the line emission in the velocity ranges indicated in the legends, while the images show the continuum emission at 218.935 GHz.  The corresponding beam sizes are shown at the bottom left corner of each panel.  The emission at the inner $\pm2$ \kms\ is excluded to avoid the kinematically quiescent gas close to the source velocity. The green ``+'' marks the ALMA continuum peak, while the cyan ``+'' marks the location of a hot core candidate to the southwest of the source. Two protostars in the field are marked with black ``+''s.}
    \label{fig:other_alma}
\end{figure*}

The $^{13}$CO, C$^{18}$O, H$_2$CO and CH$_3$OH ALMA data is available via 12-m and 7m-array observations, which we have combined. The blue and redshifted emission from these tracers is presented in Figure~\ref{fig:other_alma}. These CO maps also show extended redshifted emission to the NE and blueshifted emission to the SW. However, with the higher angular resolution, we also identify the presence of a hot core 2.13\arcsec\ to the SW of the G11.94 protostar (marked by a cyan ``+'' in Fig.~\ref{fig:other_alma}).

\begin{figure*}
    \centering
    \includegraphics[width=0.75\textwidth]{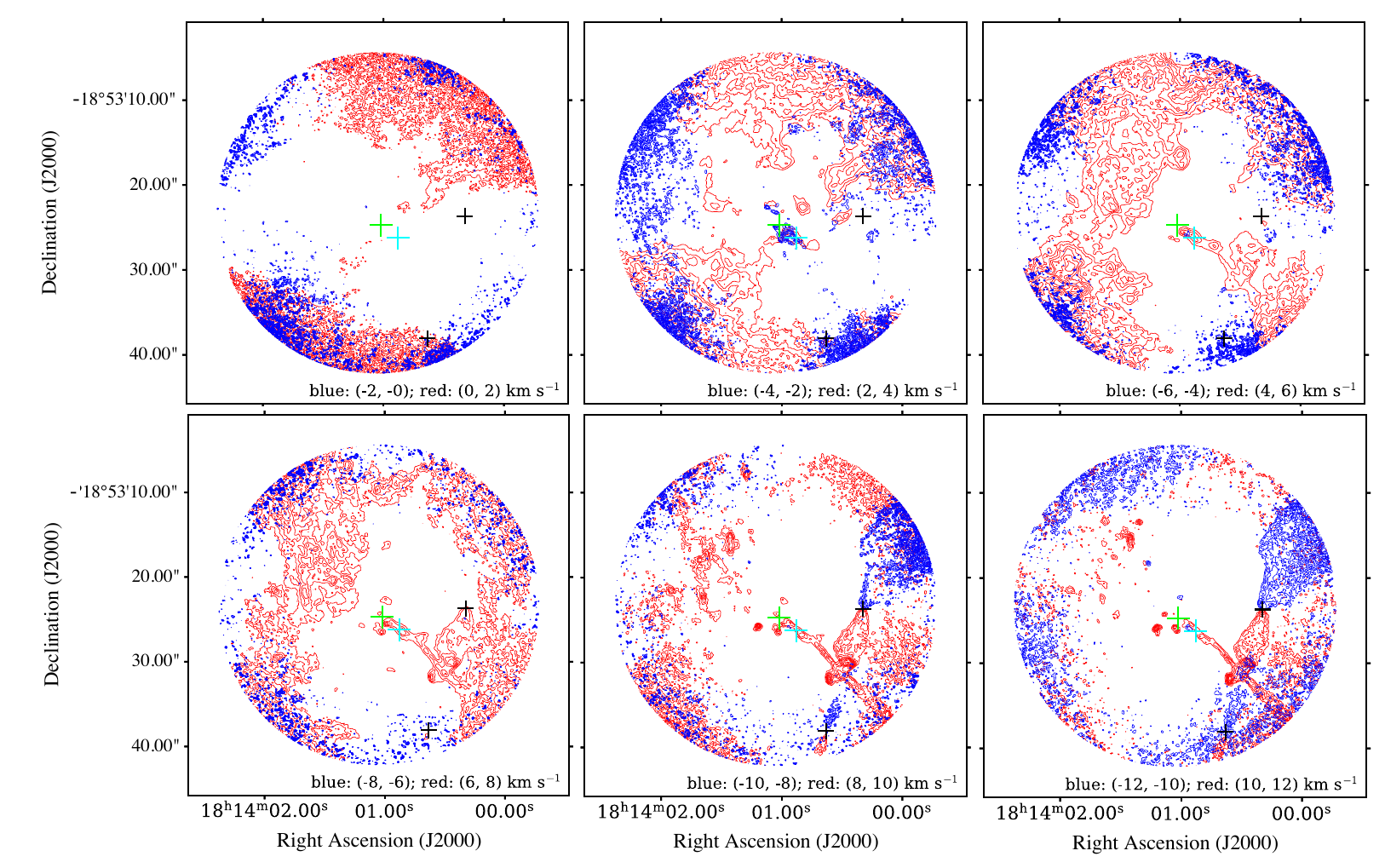}
    \caption{Channel maps of $^{12}$CO emission in 2 \kms\ intervals at both the red- and blue-shifted velocities.  The green ``+'' marks the ALMA continuum peak, while the cyan ``+'' marks the location of a hot core candidate to the southwest of the source. The two protostars in the field (black ``+'') can both be seen driving outflows with both blue lobes to the NNE and both red lobes to the SSW. The peculiar red-shifted filament emanating from G11.94 extending to the southwest can be seen above 6 \kms, in the same direction as the blue-shifted \OIII\ and \OI\ emission detected with FIFI-LS.}
    \label{fig:12co_alma}
\end{figure*}

From the high resolution $^{12}$CO data set (Fig.\,\ref{fig:12co_alma}), we also identify two nearby protostars in the G11.94 field (marked with black ``+''s in Figs. \ref{fig:multipanel}, \ref{fig:other_alma}, and \ref{fig:12co_alma}) region, one to the west and one to the south. These are detected through the presence of both compact continuum emission as well as clear bipolar $^{12}$CO outflows, with the blue-shifted lobe extending to the NNW and the red-shifted lobe extending to the SSE in both outflows. The molecular emission from these protostars contaminates the analysis of the outflow morphology from G11.94, particularly in the SW lobe implied from the \OI\ and \OIII\ morphology. 

In summary, we detect prominent redshifted emission to the northeast of G11.94 in $^{12}$CO (both combined 12-m+7-m and TP), $^{13}$CO, C$^{18}$O, H$_2$CO, and CH$_3$OH. The $^{12}$CO emission in particular possesses relatively high velocities (5-12 \kms), including several compact spots at $\sim$10 \kms. At low velocity ($\lesssim5$ \kms), the $^{12}$CO emission also appears in the northwest and southeast of G11.94, perpendicular to the inferred outflow direction, likely tracing the gas in the protostellar envelope (Fig. \ref{fig:12co_alma}).  To the southwest there is clear, collimated blue-shifted emission emanating from the continuum peak in both $^{13}$CO and C$^{18}$O, and such a morphology is consistent with the emission of H$_2$CO and CH$_3$OH. In the latter two tracers, it is possible that some of the blue-shifted emission is due to the nearby protostar, although no corresponding red-shifted emission is detected. We therefore associate it with an outflow from G11.94 given its coincident spatial correspondence with emission in $^{13}$CO and C$^{18}$O. This blue-shifted emission is not clearly detected in the combined 12-m+7-m $^{12}$CO data, but is evident in the $^{12}$CO TP image. 

Interestingly, in the high resolution $^{12}$CO data, there is a narrow, high-velocity, red-shifted filament extending to the SW.  The nature of this filament is puzzling, as it extends in the same direction as the inferred blue-shifted outflow from FIFI-LS, as well as the $^{13}$CO and C$^{18}$O lines, and in the opposite direction as the higher-velocity red-shifted emission seen across all CO isotopologues. There is a velocity gradient in this filament: the emission increases in velocity with distance from either G11.94 or the hot core candidate. This filamentary structure may be interacting with the outflow from the nearby protostar to the NW, as suggested by the overlapping $^{12}$CO emission. We also see greater emission of H$_2$CO and CH$_3$OH near the overlapped $^{12}$CO emission.  Further analysis is required to fully understand the nature of this filament, as well as its possible interaction with another protostellar outflow.

Finally, we examine the NIR data obtained with the LBT for Br$\gamma$ and H$_2$ lines and K band continuum. An RGB image of these three images is presented in Figure~\ref{fig:lbt}). These LBT data reveal a large knot of H$_2$ emission (red in Fig. \ref{fig:lbt})  approximately 1.5\arcmin\ to the southwest of G11.94, along with two smaller knots of emission to the southeast and extended faint H$_2$ emission to the northwest. Br$\gamma$ emission largely traces the emission structures observed in the NIR and MIR (see Figure \ref{fig:multipanel}). We also note the presence of a dust lane, oriented on an approximately N-S axis, which bisects and obscures part of the region immediately to the east of the source.

\subsection{Discussion on Outflow Orientation}

While the \OI\ and \OIII\ emission may hint at a simple outflow morphology, the ALMA data depict a more complex picture.  There has yet to be a comprehensive census of the outflow morphology of this region. An outflow from G11.94 was already implied from the broad \NeII\ and $^{12}$CO line width \citep{1996ApJ...457..267S,zhu_ne_2008}, and this is confirmed via the detection of broad \OIII\ and \OI\ lines presented in this work. Here we discuss two proposed scenarios for the nature of the outflow from G11.94.

\begin{figure}[htbp!]
    \centering
    \includegraphics[width=0.47\textwidth]{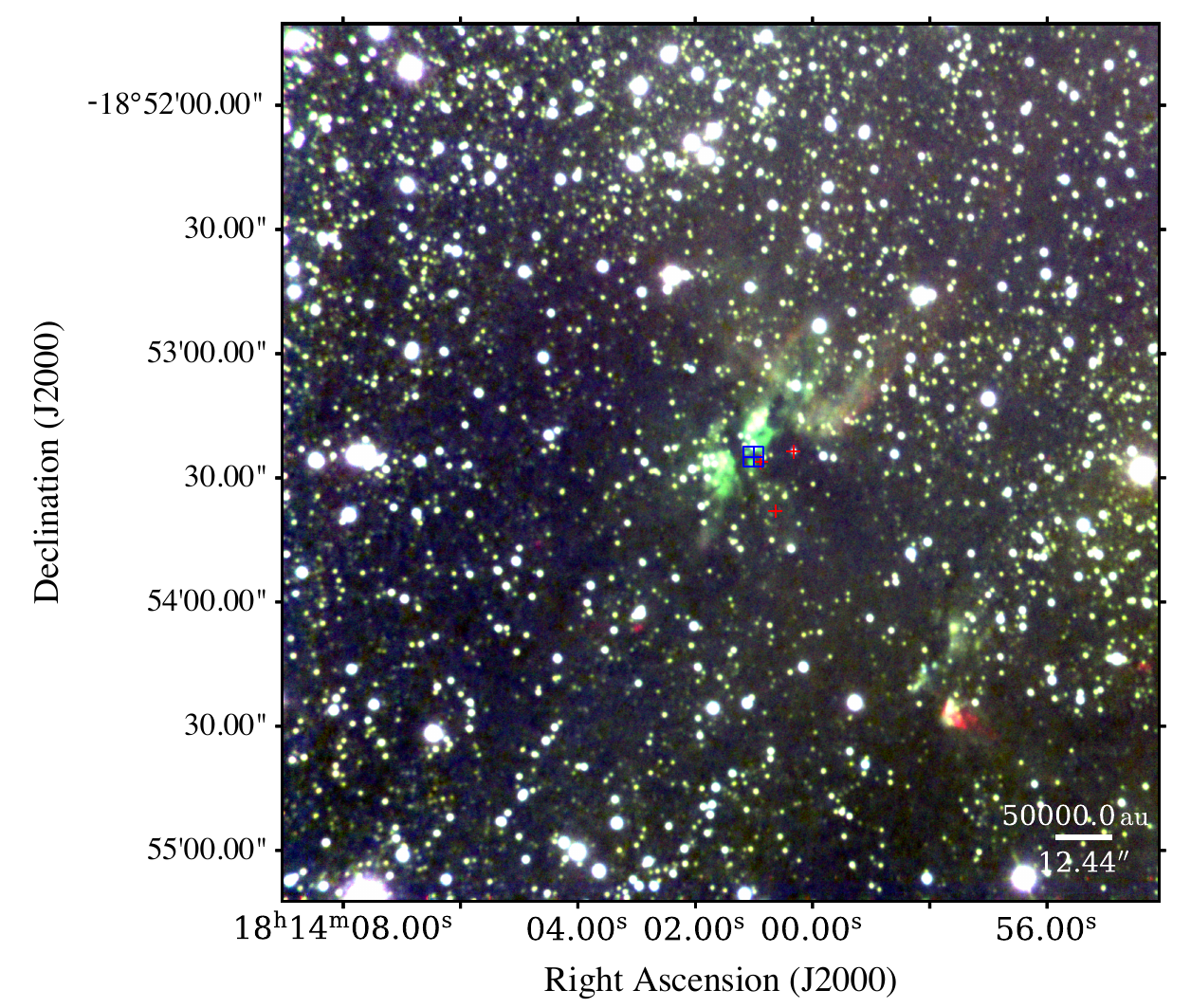}
    \caption{NIR imaging of G11.94 with the LBT. Red channel is H$_2$, green is Br$\gamma$, and blue is K band continuum. The source is identified with a magenta cross, and nearby protostars are shown with cyan. The red circle highlights a large H$_2$ knot approximately 1.5\arcmin\ to the SW of G11.94.}
    \label{fig:lbt}
\end{figure}

From the observations of FIFI-LS and ALMA, an outflow axis that is oriented in a northeast-southwest direction is implied, with the red lobe extending to the northeast and the blue lobe extending to the southwest.  The morphology of \OI, \OIII, and CO isotopologues indicates a wide-angle outflow, especially in the northeastern lobe. While the blue-shifted southwestern lobe is not clearly seen in $^{12}$CO, except for the channels between $(-6, -2)$ $\kms$ (Figure\,\ref{fig:12co_alma}), possibly due to confusion with quiescent gas, the blue-shifted lobe can be traced by the $^{13}$CO and C$^{18}$O lines, which trace denser gas, as well as the emission of methanol and formaldehyde (Figure\,\ref{fig:other_alma}).  We estimate a $\sim100^\circ-110^\circ$ for the edge-to-edge outflow opening angle of the red-shifted lobe. Such a large outflow opening angle can result in a flattened structure which appears to be perpendicular to the outflow direction \citep{2014ApJ...788..166Z}, similar to the morphology of the low-velocity $^{12}$CO emission. This could explain some of the perpendicularly elongated MIR and NIR emission features seen in Figures \ref{fig:multipanel} and \ref{fig:lbt}, respectively. The blue-shifted lobe appears to have a narrower opening angle than that of the red-shifted lobe. The large H$_2$ knot to the SW detected by the LBT (Figure\,\ref{fig:lbt}) is aligned well with the blue-shifted outflow axis suggested by this interpretation.

It is expected that massive protostars begin to ionize their surroundings in the later stages of their evolution as they contract towards the ZAMS. Moreover, the outflow opening angle is expected to widen significantly as protostars evolve  \citep[e.g.,][]{2005ASSL..324..105B,2007prpl.conf..245A,2014prpl.conf..149T}. Thus, a region with ionized emission could also possess a wide-angle outflow. The presence of \OIII\ emission in G11.94, together with the morphology of molecular lines observed with ALMA, supports this scenario. 

However, there are some issues with the interpretation of a NE-SW outflow. Both the near-IR continuum observed with Spitzer IRAC bands as well as the Br-$\alpha$ line imaged by the LBT show extended emission in the northwest-southeast directions, perpendicular to the northeast-southwest outflow discussed in previous paragraphs.  This emission in the NW-SE direction could be an outflow. However, there are several reasons why we favor the outflow in the NE-SW direction instead of this scenario.  First, the morphology of the near-IR continuum and the Br-$\alpha$ line does not necessarily contradict the NE-SW outflow axis.  In the outflow with a wide opening angle, the near-IR radiation can be mostly scattered off a wide cavity, resulting in emission elongated perpendicular to the outflow axis.  Another issue with this NW-SE outflow scenario is that only low velocity ($\lesssim5$ \kms) $^{12}$CO emission traces gas at the same NW-SE direction, which is toward the edge of the ALMA FoV with no apparent connection to G11.94 or any other continuum sources.  Compared to the broad line width in \OI\ and \OIII, this low-velocity CO emission is unlikely to trace the same material; thus, this does not strongly support the NW-SE outflow scenario. 

The faint extended H$_2$ emission to the NW seen with the LBT may support a NW-SE interpretation. However, closer investigation shows that it may be separated into two components. The western component is consistent with emission from the nearby protostar to the west, and the eastern component is consistent with a cavity wall from a wide-angle outflow emanating from G11.94. 

One feature that the NE-SW outflow, traced by \OI\ and \OIII, cannot explain is the high-velocity red-shifted filament at the southwest of G11.94 seen in $^{12}$CO.  We speculate that this filament may be a jet from the hot core candidate. 

In summary, we conclude that the most likely scenario is that the \OI\ and \OIII\ emission is tracing an outflow in the NE-SW direction, corroborated by the emission of CO isotopologues, methanol, and formaldehyde.  This outflow is poorly collimated with a wide opening angle.  The red-shifted lobe faces the northeast, while the blue-shifted lobe faces the southwest. This scenario is the most consistent with an evolved protostar driving the ionization at the base of its outflow cavity.

\subsection{Outflow Orientation from FIFI-LS Data}
\label{sec:orientation_FIFI}

The FIFI-LS continuum and line emission morphology provide three independent measurements of the outflow axis orientation. The position angle (PA) can be determined from \OIII\ and \OI\ $^3P_2\rightarrow^3P_1$ at 63 \micron\ emission (PA$_\text{\OIII}$ and PA$_\text{\OI}$). As a result of the relatively low spatial resolution, the emission features can be approximated as 2D Gaussians.  We fit 2D Gaussian peak positions of the blue- and red-shifted emission and measured the PA from these two positions.  The PA$_\text{\OIII}$ is $-144.0\pm15.0\degree$ and the PA$_\text{\OI}$ is $-141.7\pm11.2\degree$.

We can also measure the continuum peak offsets from the red to the blue channel of FIFI-LS.  The PA$_\text{cont}$ discussed in \S\ref{sec:continuum}, $-126.2\pm12.4\degree$ and $-88.5\pm19.3\degree$ are the two measurements (from 52 and 63 \micron, respectively) for the direction of the blue-shifted lobe. However, given the small magnitude of the offset for the latter measurement ($1.0\pm0.3\arcsec$), we should treat the first measurement (with an offset of $2.5\pm0.5\arcsec$) as more robust. The agreement between the position angles calculated from FIFI-LS \OI\ and \OIII\ emission indicates that the position angle for this suggested outflow is $\sim-142\degree$. Furthermore, the PA derived from oxygen lines is consistent with the PA calculated for the relative offset between 52 \micron\ and 186 \micron\ continuum imaging. 

\subsection{Mass and Momentum Outflow Rates}\label{sec:outflowrate}

\subsubsection{The Ionized Outflow}
\label{sec:ionized_outflowrate}

Using the flux of the \OIII\ line in G11.94, we attempt to infer the oxygen column density and place limits on the outflow rates of the ionized outflow.  We relate the line luminosity of \OIII\ at 52 \micron\ to the gas mass traced by \OIII.  First, the line luminosity can be expressed as a volume integral:
\begin{equation}
    L_{[\rm O\,\textsc{III}]}= \int_V f_V n_e n({\rm O}^{2+})j_{[\rm O\,\textsc{III}]} (n_e, T_e) \mathrm{d}V,
    \label{eq:L_OIII}
\end{equation}
where $f_V$ is the volume filling factor, $n({\rm O}^{2+})$ is the density of ionized oxygen, $j_{[\rm O\,\textsc{III}]} (n_e, T_e)$ is the emissivity of the \OIII\ 52 \micron\ line, which is a function of electron density ($n_e$) and electron temperature ($T_e$).  We assume that within the ionized region, which is treated as being distinct from the atomic part of the outflow traced by \OI, all O is in the $\rm O^{2+}$ ionization state and adopt an O abundance of $X_{\rm O}=3\times10^{-4}$ with respect to H nuclei. Within the ionized region we also assume that He is singly ionized, so that $n_e=1.1 n_{\rm H}$.  Then the 52~$\rm \mu m$ line flux can be expressed as
\begin{equation}
    F_{[\rm O\,\textsc{III}]} = \frac{L_{[\rm O\,\textsc{III}]}}{4\pi d^2} = \frac{f_V V}{1.1}\langle n_e^2 \rangle X_{\rm O} \frac{j_{[\rm O\,\textsc{III}]} (n_e, T_e)}{4\pi d^2},
    \label{eq:F_OIII}
\end{equation}
where $d$ is the distance to the source.  

The average oxygen column density can be evaluated via a volume integral as 
\begin{equation}
    N_{\rm O} = \frac{1}{\Omega_\text{aper}d^2}\int_V f_V n_{\rm O} \mathrm{d}V = f_V X_{\rm O} \frac{\langle n_e \rangle V}{1.1 \Omega_\text{aper}d^2},
    \label{eq:N_OIII}
\end{equation}
where $\Omega_\text{aper}$ is the solid angle of the aperture.  If we assume an uniform distribution so that $\langle n_e^2 \rangle = \langle n_e\rangle^2$ and combine Equations\,\ref{eq:F_OIII} and \ref{eq:N_OIII}, we obtain
\begin{eqnarray}
    F_{[\rm O\,\textsc{III}]} & = & N_{\rm O} \frac{\Omega_\text{aper}}{4 \pi} \langle n_e \rangle j_{[\rm O\,\textsc{III}]} (n_e, T_e),\\
    & = & N_{\rm H} X_{\rm O}\frac{\Omega_\text{aper}}{4 \pi} \langle n_e \rangle j_{[\rm O\,\textsc{III}]} (n_e, T_e),
\end{eqnarray}
where $N_{\rm O}$ and $N_{\rm H}$ are the column densities of O and H nuclei that are associated with the \OIII\ emitting, i.e., ionized, region.

\begin{figure}[htbp!]
    \centering
    \includegraphics[width=\linewidth]{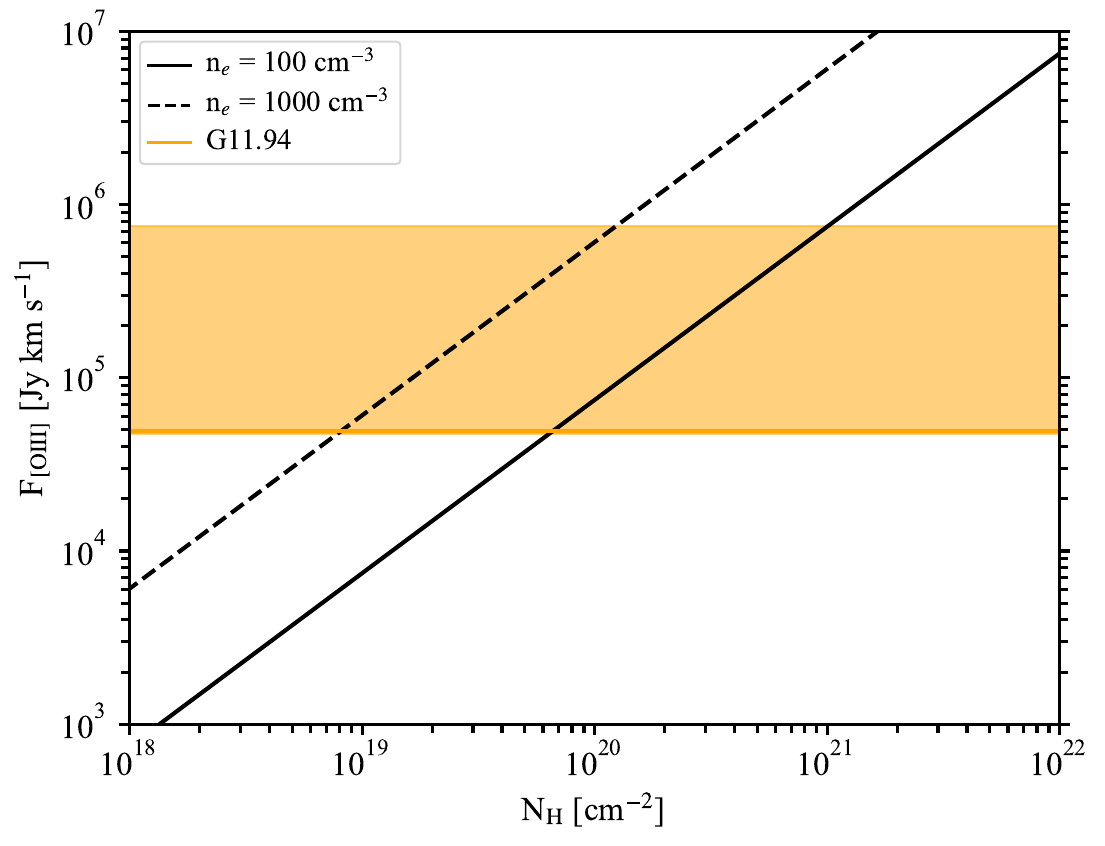}
    \caption{Expected flux of the \OIII\ 52 \micron\ as a function of column density of H nuclei in the ionized region, which is assumed to extend over a 10\arcsec\ radius aperture, based on \texttt{PyNeb} models for a temperature of $10^4\:$K and densities of $n_e=100\:{\rm cm}^{-3}$ (black solid line) and $1000$ $\:{\rm cm}^{-3}$ (black dashed line).  The observed flux from G11.94 is shown by the horizontal yellow line. The horizontal shaded band shows a range of flux values extending upwards by a factor of 14.9 that accounts for potential effects of dust extinction (see text).}
    \label{fig:pyneb_oiii}
\end{figure}

Using \texttt{PyNeb} \citep{2015A&A...573A..42L}, we calculate the emissivity of \OIII\ at given values of $n_e$ and $T_e$ and thus link the observed flux of the \OIII\ line to the column of the ionized gas. Figure~\ref{fig:pyneb_oiii} shows example model results for cases with $n_e=100$ and 1000 cm$^{-3}$, both with $T_e = 10^4$ K, with the extent of the ionized region guided by the observed size of \OIII\ emission from G11.94, i.e., a radial extent of 10\arcsec, equivalent to $R_{\rm out}=0.2\:$pc (or 40,000~au). We see below that these values of density are representative of self-consistent solutions for the emission from an ionized outflow from G11.94.

When comparing to the observed \OIII\ flux, we need to consider the impact of dust extinction.  As a fiducial value for the mass surface density of the extincting material, we adopt 50\% of that of the clump environment, i.e., $\Sigma=\Sigma_{\rm cl}/2=0.87\:{\rm g\:cm}^{-2}$, which is representative of the value inferred from SED model fitting.  The value of the opacity at 52~$\rm \mu m$ depends on the adopted dust model.  For the dust model of \citet{1994ApJ...422..164K}, i.e., the dust model used in the radiative transfer simulations of \citet{2018ApJ...853...18Z}, the value of $\kappa$ at 52 \micron\ is $1.084\:{\rm cm^2\:g}^{-1}$ (note, per gram of gas). However, for the moderately coagulated thin ice mantle model of \citet{1994AA...291..943O} (hereafter the OH5 dust model), the equivalent $\kappa$ is $3.09\:{\rm cm^2\:g}^{-1}$ (per gram of gas). Therefore we consider that the optical depth of the foreground material could be in the range $\tau=\kappa\Sigma\simeq 0.9 - 2.7$ so that the intrinsic flux of \OIII\ emission may be 2.5 to 14.9 times greater than the observed value. Given these considerations, in Fig.~\ref{fig:pyneb_oiii} we show a range of values for the \OIII\ flux from G11.94, from the observed value of $\sim5\times10^4\:{\rm Jy\:km\:s}^{-1}$ up to $2.3\times10^5\:{\rm Jy\:km\:s}^{-1}$, and adopt a fiducial value of $1\times10^5\:{\rm Jy\:km\:s}^{-1}$.

With the above results, we estimate that the average column density of the ionized outflow is $N_{\rm H}\sim 2\times10^{19}\:{\rm cm}^{-2}$ for the cases with $n_e=1000\:{\rm cm}^{-3}$ and $2\times10^{20}\:{\rm cm}^{-2}$ for the case with $n_e=100\:{\rm cm}^{-3}$, respectively. Given the morphology of the \OIII\ emission we expect a volume filling factor of order unity and thus adopt $f_V=1/3$ in a spherical volume of 10\arcsec\ radius. This implies a gas density in the ionized outflow of $n_{\rm H}= 3 N_{\rm H} / (4 f_V R_{\rm out})\rightarrow 75$ and $750\:{\rm cm}^{-3}$. Thus, approximately self-consistent results are achieved for the case $n_e=300\:{\rm cm}^{-3}$, which implies $N_{\rm H}\simeq8.7\times10^{19}\:{\rm cm}^{-2}$ and $n_{\rm H}\simeq 325\:{\rm cm}^{-3}$, i.e., $n_e\simeq 360\:{\rm cm}^{-3}$. 

With this value of $n_{\rm H}=325\:{\rm cm}^{-3}$ or $N_{\rm H}=8.7\times 10^{19}\:{\rm cm^{-2}}$, we estimate a mass of the ionized outflow of $M_{\rm out,OIII}=0.12\:M_\odot$ that is present within a 0.2~pc radius of the source. To calculate the mass flux in this component of the outflow, we estimate the timescale for the outflow to cross this region, i.e., to flow from the protostar to a radial distance of 0.2~pc. For this, we adopt an outflow speed of $v_{\rm out,OIII}=$200~\kms, i.e., the observed HWHM of the \OIII\ emission at 52~\micron\ with FIFI-LS. Given the uncertainty in the orientation of the source, for simplicity we assume this is representative of the 3D speed of the outflow.  Thus the outflow timescale is $t_{\rm out} = R_{\rm out}/v_{\rm out}= 980\:$yr. Thus the mass flux in the \OIII-traced outflow is $1.2\times 10^{-4}\:M_\odot\:{\rm yr}^{-1}$. Similarly, the total momentum in the ionized component of the outflow is $p_{\rm out,OIII}=m_{\rm out,OIII}v_{\rm out,OIII} = 16\:M_\odot\:{\rm km\:s^{-1}}$ and the momentum outflow rate is $\dot{p}_{\rm out,OIII}=0.025\:M_\odot\:{\rm km\:s^{-1}\:yr^{-1}}$. These results are listed in Table \ref{tab:outflow_properties}.

We note that the protostellar accretion rate estimated from SED fitting has a value of $\sim9\times10^{-4}\:M_\odot\:{\rm yr}^{-1}$. The primary outflow is thus expected to have a mass injection rate of about 10\% of this \citep[e.g.,][]{2018ApJ...853...18Z}, i.e., $\sim 9\times10^{-5}\:M_\odot\:{\rm yr}^{-1}$. We see that the ionized component of the outflow has a mass flux that is very similar to that expected for the total primary outflow.

We also note that \OIII\ emission from massive protostars is relatively rare, with G11.94 being one of the few SOMA Atomic Outflow Survey sources that shows such strong emission. Recently, \citet{2025A&A...697A.186K} have reported \OIII\ emission from the massive protostar DR21 Main, with their set-up including both the 52 and 88~$\rm \mu m$ lines. These authors also used \texttt{PyNeb} to infer conditions in the ionized gas, deriving $T_e \sim 8000$ K and $n_e \sim 250$ cm$^{-3}$, i.e., similar densities to those we have inferred.

\begin{table*}
    \centering
    \begin{tabular*}{0.71\textwidth}{c|c|c|c|c|c|c}
        \hline
        \hline
        & $m_{\rm out}$ & $v_{\rm out}$ & $p_{\rm out}$ & $t_{\rm out}$ & $\dot{m}_w$  & $\dot{p}_{w}$ \\
        & $(M_\odot)$ & ($\rm km\:s^{-1}$) & ($M_\odot$ \kms) & (yr) & ($M_\odot$ yr$^{-1}$) & ($M_\odot$ \kms yr$^{-1}$) \\
        \hline  
        Ionized & 0.116 & 211 & 24.5 & 980 & $1.2\times10^{-4}$ & $2.5\times10^{-2}$ \\
        \hline  
        Atomic & 0.038 & 43.3 & 1.65 & 7,500 & $5.1\times 10^{-6}$ & $2.2\times10^{-4}$ \\
        \hline
        Molecular & 4.14 & 25.9 & 107 & 15,700 & $2.6\times10^{-4}$ & $6.8\times10^{-3}$ \\
        \hline
        \hline
    \end{tabular*}
    \caption{Ionized, atomic and molecular outflow properties of the massive protostar G11.94 (see text).}
    \label{tab:outflow_properties}
\end{table*}

\subsubsection{The Atomic Outflow}
\label{sec:atomic_outflowrate}

We model the \OI\ line intensities and line ratios to estimate the column density of material associated with this emission and therefore the mass and momentum flux of this ``atomic'' component of the G11.94 outflow.  However, the \OI\ 63 \micron\ line is known to have self-absorption at line center \citep{2015A&A...584A..70L}.  Thus, simply dividing the peak intensity of the two \OI\ lines is unlikely to provide a realistic measurement of the line ratio.  To avoid the absorption-contaminated line center, we estimate the line ratio on the wings of these two lines, although the 63 \micron\ \OI\ fit is further hindered by the large telluric feature on the red side of the spectrum.  The two \OI\ lines were observed with different instrumental spectral resolution (\S\ref{sec:obs_fifils}).  We first convolve the \OI\ 63 \micron\ spectrum to match the spectral resolution at 145 \micron.  Then, we divide the spectra of the 63 \micron\ over the \OI\ 145 \micron\ and take the average across -300 to -100 \kms, exclusively on the blue portions of the spectra.  The measured ratio is $2.30\pm0.20$.

However, again, dust extinction is another factor that may affect the relative intensity of the \OI\ 63 and 145 \micron\ lines. Specifically, the higher opacity at 63 \micron\ preferentially lowers the luminosity of the \OI\ 63 \micron\ line. The \citet[][]{1994ApJ...422..164K} dust model $\kappa$ values at 63 \micron\ and 145 \micron\ are 0.735 and 0.133 cm$^2$ g$^{-1}$, respectively. If the OH5 dust model is used, then these values would be 2.09 and $0.47\:{\rm cm^2\:g}^{-1}$, respectively. The ratio of the \OI\ line intensities is affected by extinction via:
\begin{equation}
    \frac{I_{\rm 63,obs}}{I_{\rm 145,obs}} = \frac{I_{\rm 63,true} e^{-\tau_{63}}}{I_{\rm 145,true}e^{-\tau_{145}}} = \frac{I_{\rm 63,true}}{I_{\rm 145,true}}e^{-(\Delta\kappa_{63-145}\Sigma)},
\end{equation}
where $\Delta \kappa_{63-145}=\kappa_{\rm 63\mu m}-\kappa_{\rm 145\mu m}$. Again we adopt an extincting mass surface density of $\Sigma=\Sigma_{\rm cl}/2=0.87\:{\rm g\:cm}^{-2}$. Under such conditions, $e^{-(\Delta\kappa_{63-145}\Sigma)}=0.59$ for the \citet{1994ApJ...422..164K} dust model and 0.24 for the OH5 dust model. Given these considerations, we consider a range of values of line intensity ratios from 2.3 to 9.6.

For our modeling of expected line ratios and intensities, we utilize a 1-D adaptation of the {\sc 3d-pdr} software \citep{2012MNRAS.427.2100B}, considering a range of uniform densities, ${\rm log}(n_{\rm H}/{\rm cm}^{-3})=2,3,4,5$, and a range of UV flux densities, log($\chi/\chi_0)=1,3,5$ \citep[normalized to the radiation field of][]{1978ApJS...36..595D}. We assume a constant cosmic ray ionization rate of $10^{-16}$ s$^{-1}$ \citep{2006PNAS..10312269D,2022A&A...658A.151G}, solar metallicity composition, and ``turbulent'' (i.e., kinematic) broadening of 100~km~s$^{-1}$, which is designed to reflect the velocity distribution of the outflow. We model the \OI\ fluxes at 63 and 145 \micron\ in each simulated PDR region by solving the respective radiative transfer equation \citep{2017ApJ...850...23B}. 

\begin{figure*}[htbp!]
    \centering
    \includegraphics[width=\textwidth]{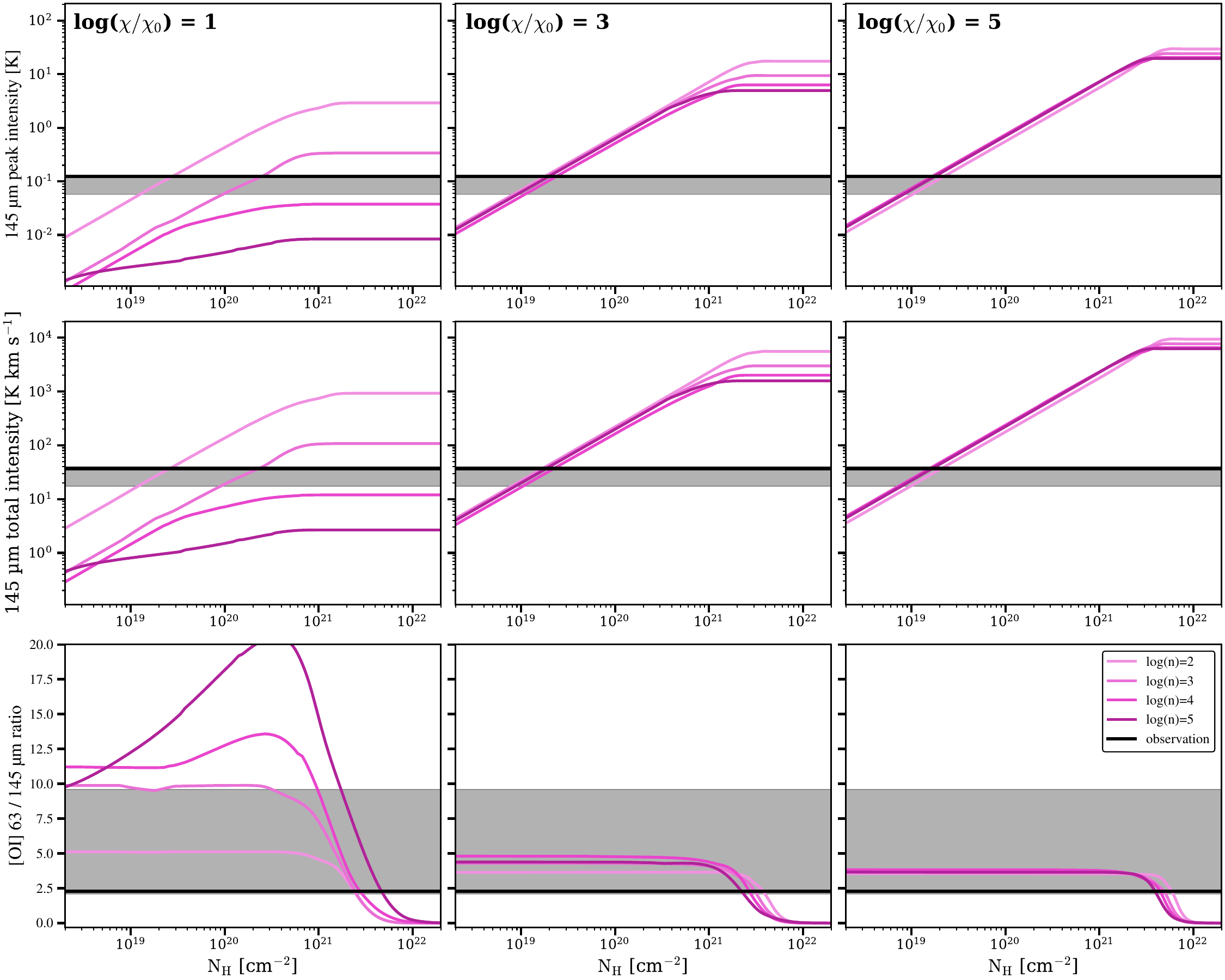}    
    \caption{{\sc 3d-pdr} modelling to predict \OI\ emission properties from total gas column density, $N_{\rm H}$, for a range of FUV flux densities (columns) and a range of environment densities (see legend). From top to bottom, gas column density is plotted against \OI\ 145 \micron\ peak intensity, \OI\ 145 \micron\ total intensity, and \OI\ 63/145 \micron\ ratio. Uncertainties are derived from combining the uncertainty from flux measurements (negligible in rows 1 and 2) with the range of possible dust extinction, which dominates the uncertainty in all plots. The flux ratio measurement in the third row is more uncertain because the ratio is calculated from a line wing fit.}
    \label{fig:PDR_models}
\end{figure*}

To obtain an estimate for the total column density of H nuclei, $N_{\rm H}$, we use three metrics: the peak intensity of the 145 \micron\ line ($I_{\rm 145,peak}$); the total flux of the 145 \micron\ line ($F_{145}$); and the ratio of the 63 \micron\ to 145 \micron\ line intensities ($I_{\rm 63}/I_{145}$). We note that the peak intensity is impacted by the instrumental resolution and we convolve the model with the appropriate value at each wavelength. From the results shown in Figure \ref{fig:PDR_models} we conclude that for the higher FUV field cases that we consider to be most relevant, values of $N_{\rm H}\simeq 2 \times 10^{19}\:{\rm cm}^{-2}$ are favored.  Given the uncertainties in the line ratio induced by potential effects of differential extinction, this ratio provides little constraint on the volume density of the gas.

Following the methodology from Section \ref{sec:ionized_outflowrate} and adopting a radius of 17\arcsec\ for the extent of the \OI\ outflow, (68,000~au, $R_{\rm out,OI}=0.33\:$pc), the mass in the \OI\ outflow is estimated to be $m_{\rm out,OI}=0.038\:M_\odot$. To calculate the mass flux in this component of the outflow we estimate the outflow crossing timescale, i.e., flow from the protostar to a radial distance of 0.33~pc. For this we adopt a line-of-sight outflow speed of 43.3~\kms, i.e., the observed HWHM of the \OI\ emission at 145~\micron. As for the ionized outflow traced by \OIII, we also assume that the intrinsic 3D velocity is well represented by this value.  Thus the outflow timescale is $t_{\rm out,OI} = R_{\rm out,OI}/v_{\rm out,OI}= 7,500\:$yr and the implied mass flux is $\dot{m}_{\rm out,OI}=5.1\times10^{-6}\:M_\odot\:{\rm yr}^{-1}$. The momentum in the observed outflow aperture is estimated to be $p_{\rm out,OI}=m_{\rm out,OI}v_{\rm out,OI}=1.6\:M_\odot\:{\rm km\:s}^{-1}$ and the momentum outflow rate is $\dot{p}_{\rm out,OI}=2.2\times 10^{-4}\:M_\odot\:{\rm km\:s^{-1}\:yr}^{-1}$.

Comparing again to the protostellar accretion rate estimated from SED fitting, which has a value of $\sim9\times10^{-4}\:M_\odot\:{\rm yr}^{-1}$, and thus the expected primary outflow mass injection rate of $\sim 9\times10^{-5}\:M_\odot\:{\rm yr}^{-1}$, we see that the atomic component of the outflow is $\lesssim10\%$ of the expected total mass flux of the primary outflow.


\subsubsection{The Molecular Outflow}
\label{sec:tp}

To estimate mass outflow rate of the molecular outflow, we utilize the ALMA Total Power (TP) observations of the $^{12}$CO 2$\rightarrow$1 line, which have an angular resolution (half-power beam width) of 22.6\arcsec. We extract a spectrum from a beam area centered on the protostar, i.e., with radial extent of 22.6\arcsec, i.e., 90,000~au, i.e., 0.44~pc.  Utilizing methodology from \citet{2014ApJ...783...29D} and \citet{2019ApJ...873...73Z}, we derive the outflow mass and momentum from the TP spectral cube, excluding the inner $\pm5$ \kms\ to avoid confusion with any surrounding quiescent gas. We assume optically thin emission, a $^{12}$CO abundance of $10^{-4}$ with respect to H$_2$ \citep{1982ApJ...262..590F, 1994ApJ...428L..69L, 2007A&A...472..187H} and an excitation temperature $T_{\text{ex}}$ for the $^{12}$CO ($2\rightarrow1$) transition of 17.5 K. This temperature provides the minimum estimate for outflow mass.  Picking a $T_{\text{ex}}$ of 50.0 K, which is generally the upper limit for mass estimates using low-$J$ CO transitions, would increase our estimates for mass by a factor of 1.5. We calculate the outflow mass in each velocity channel and then sum over the spectrum to obtain a total CO-traced outflow mass of $m_{\rm out,mol}=4.14\:M_\odot$. We evaluate the mass-weighted mean line-of-sight velocity to be $25.9\:{\rm km\:s}^{-1}$ and, again, assume this representative of the 3D velocity.  The total momentum is then estimated to be $107\:M_\odot\:{\rm km\:s}^{-1}$. The outflow timescale is $t_{\rm out,mol}=R_{\rm out,mol}/v_{\rm out,mol}=15,700\:$yr. Thus the mass and momentum fluxes in the molecular outflow are $\dot{m}_{\rm out,mol} = 2.6\times 10^{-4}\:M_\odot\:{\rm yr}^{-1}$ and $\dot{p}_{\rm out,mol}=6.8\times10^{-3}\:M_\odot\:{\rm km\:s^{-1}\:yr^{-1}}$.

Comparing to the other components, we see that the molecular outflow appears to dominate the total mass flux, i.e., being 3.1 times greater than the ionized component. However, the momentum flux of the molecular outflow is about 2.5 times smaller than that of the ionized component. The atomic outflow is a very minor component in both metrics, between $\sim1-10\%$ that of the molecular and ionized components. These results imply that the ionized outflow dominates the contribution of the total outflow force, albeit with a significant fraction, i.e., $\sim 30\%$, being carried by molecular gas. The relatively low velocity of the molecular gas and its comparatively high mass flux rate suggest that it is primarily composed of swept-up gas from the natal core envelope.

\section{Conclusions}
\label{sec:conclusion}

We have presented first results from the SOMA Atomic Outflow Survey, which has used SOFIA/FIFI-LS to observe a sample of massive protostars, in the \OI\ lines at 63 and 145 \micron, the ionized \OIII\ line at 52 \micron, and the CO line at 186 \micron. Here we have focused on the source G11.94, which presents relative strong \OIII\ emission.  The main conclusions of our investigation are as follows:
\begin{enumerate}[nosep]
    \item The massive protostar G11.94 exhibits strong ionized and atomic oxygen lines with broad line widths: $\sim200$ \kms\ in \OIII\ and $\sim40-80$ \kms\ in \OI.
    \item The kinematic distribution of the \OIII\ and \OI\ emitting gas suggests a bipolar outflow with the red lobe extending to the northeast and the blue to the southwest. We infer a position angle for the blue-shifted lobe of approximately $-143\degree$.
    \item This ionized and atomic outflow is corroborated by molecular emission in ALMA observations of $^{12}$CO, $^{13}$CO, C$^{18}$O, H$_2$CO, and CH$_3$OH. From these observations, it appears that the blue-shifted lobe of the molecular outflow is more collimated than the brighter, wide ($\sim110\degree$) angle red-shifted outflow lobe.
    \item LBT imaging in the NIR reveals a complicated morphology. One interpretation is that NIR-bright emission traces the base of wide-angle outflow cavities. 
    \item We have estimated a mass flux of the ionized outflow that is about 10\% of accretion rate derived from SED-fitted Turbulent Core Accretion models of the protostar. This suggests that primary outflow, i.e., launched directly from the disk, is dominated by the ionized outflow. This is consistent with our estimates of the momentum flux of the ionized outflow, which dominates over the atomic and molecular components. However, the molecular outflow dominates the total mass flux, suggesting it is mainly composed of swept-up, secondary outflow gas. The atomic outflow is sub-dominant in both mass and momentum flux compared to the other components.
    \item A dominant ionized protostellar outflow is the expected signature of a very massive protostar in the later stages of its growth, i.e., with a protostellar structure that has contracted to near the ZAMS, leading to high rates of H ionizing luminosity. The fact that the outflow is seen in \OIII\ line emission indicates that a photospheric temperature of at least 37,000~K is required, i.e., to enable ionization of He to He$^+$, which requires a ZAMS stellar mass of at least $\sim30\:M_\odot$. Such a mass is consistent with the results of SED fitting, i.e., $m_*=22.4^{+21}_{-11}M_\odot$, although this estimate is subject to significant uncertainties. Indeed, the presence of strong \OIII\ emission then places additional constraints on these models, helping to break some of the degeneracy present in SED fitting.

    \item Future observations are need to confirm the above conclusions. In particular, improved kinematic characterization of the protostar is needed, e.g., a dynamical mass estimate from accretion disk tracers using high resolution ALMA observations. Improved kinematic characterization of the molecular, atomic and ionized outflow is also needed, e.g., with ALMA and future-FIR facilities, such as PRIMA.

\end{enumerate}

\section*{acknowledgements}
We thank an anonymous referee for help with comments. We thank the ALMA-ATOMS-QUARKS survey team for providing their fully prepared $^{12}$CO data cube. P.O. acknowledges support from a Virginia Initiative on Cosmic Origins (VICO) summer undergraduate fellowship. Y.-L.Y. acknowledges support from Grant-in-Aid from the Ministry of Education, Culture, Sports, Science, and Technology of Japan (20H05845, 20H05844, 22K20389, 25H00676), and a pioneering project in RIKEN (Evolution of Matter in the Universe). J.C.T. acknowledges support from ERC Advanced Grant MSTAR (788829) and NSF AAG grant 2206450. T.G.B. acknowledges support from the Leading Innovation and Entrepreneurship Team of Zhejiang Province of China (Grant No. 2023R01008). R.F. acknowledges support from the grants PID2023-146295NB-I00, and from the Severo Ochoa grant CEX2021-001131-S funded by MCIN/AEI/10.13039/501100011033 and by ``European Union NextGenerationEU/PRTR''.  Based in part on observations made with the NASA/DLR Stratospheric Observatory for Infrared Astronomy (SOFIA). SOFIA is jointly operated by the Universities Space Research Association, Inc. (USRA), under NASA contract NNA17BF53C, and the Deutsches SOFIA Institut (DSI) under DLR contract 50 OK 2002 to the University of Stuttgart. Additional financial support for this work was provided by NASA through award \#09\_0169 issued by USRA.

\bibliography{research}
\bibliographystyle{aasjournal}

\end{document}